\documentclass[preprint,12pt]{elsarticle}

\journal{Annals of Physics}

\RequirePackage{bm,amsfonts,lineno,hyperref,amsmath,amssymb,microtype,float}
\usepackage{color}
\begin{document}


\title{Relativistic strange stars in Tolman-Kuchowicz spacetime}

\author{Suparna Biswas$^a$, Dibyendu Shee$^b$, Saibal Ray$^{b,c}\footnote{$^*$Corresponding author.\\
{\it E-mail addresses:} sb.rs2016@physics.iiests.ac.in (SB), dibyendu\_shee@yahoo.com (DS), saibal@associates.iucaa.in (SR), rahaman@associates.iucaa.in (FR), dean.fa@iiests.ac.in (BKG).}$, F. Rahaman$^d$, B.K. Guha$^a$}

\address{$^a$Department of Physics, Indian Institute of Engineering Science and Technology, Shibpur, Howrah 711103, India\\
$^b$Department of Physics, Government College of Engineering and Ceramic Technology, Kolkata 700010, West Bengal, India \\$^c$Department of Natural Sciences, Maulana Abul Kalam Azad University of Technology, Haringhata 741249, West Bengal, India\\
$^d$Department of Mathematics, Jadavpur University, Kolkata 700032, West Bengal, India}
\date{Received: date / Accepted: date}

\maketitle

\begin{abstract}
In this article we propose a relativistic model of a static spherically symmetric anisotropic strange star with the help of Tolman-Kuchowicz (TK)
metric potentials [Tolman, Phys. Rev. {\bf55}, 364 (1939) and Kuchowicz, Acta Phys. Pol. {\bf33}, 541 (1968)]. The form of the potentials are
$\lambda(r)=\ln(1+ar^2+br^4)$ and $\nu(r)=Br^2+2\ln C$ where $a$, $b$, $B$ and $C$ are constants which we have to evaluate using boundary conditions.
We also consider the simplest form of the phenomenological MIT bag equation of state (EOS) to represent the strange quark matter (SQM) distribution
inside the stellar system. Here, the radial pressure $p_r$ relates with the density profile $\rho$ as follows, $p_r(r)=\frac{1}{3}[\rho(r)-4B_g]$,
where $B_g$ is the Bag constant. To check the physical acceptability and stability of the stellar system based on the obtained solutions, we have
performed various physical tests. It is shown that the model satisfies all the stability criteria, including nonsingular nature of the density and
pressure, implies stable nature. Here, the Bag constant for different strange star candidates are found to be $(68-70)$~MeV/{fm}$^3$ which satisfies
all the acceptability criteria and remains in the experimental range.
\end{abstract}



\section{Introduction}
Einstein's theory of General Relativity (GR)~\cite{einstein/1915} is
undoubtedly and undebatedly the most promising theory of the last
century. Though the spark of Albert Einstein was proved by his theory of Special
Relativity (SR)~\cite{einstein/1905} in front of the
entire world, whose generalisation leads us the former one. Einstein
revealed the mysteries of the universe through complex calculation of Tensor Analysis
and explained the beautiful philosophical concept that matter creates
curvature in the spacetime~\cite{Wheeler/1962}. Though, before Einstein's revolutionary discovery,
tensor analysis was considered just as a mathematical tool, not so useful in physics.
Even the concept of gravitational waves was also given by this
theoretical physicist with his mastermind thinking, at that time of minimum technology.
Einstein's field equations (EFE) for the spacetime is very essential to interpret various astrophysical
phenomena like black holes, compact objects, supernovae and the formation of structure
of the universe. After $100$ years of GR, we are now able to detect gravitational waves by
promising astronomical instruments like LIGO, LISA, Virgo etc. In a short,
his creation made a new island of perambulation for the researchers in the field
of astrophysics and cosmology.

After all the thermonuclear fuels burnt out, type II supernovae explosion occurs and the
gravitational collapse of massive stars $(M>8M_\odot)$ gives the birth of neutron stars~\cite{Wilczek/1999}.
The concept of neutron stars got concrete support with the discovery of neutron particle
in 1932 by Chadwick. Later, observationally detection of pulsars~\cite{hewish/1968}
strongly re-establish this concept. Being mostly dominated by neutron particle, neutron
stars are extensively dense and even can distort the geometry of the spacetime. Their
expanse is not very big, radius lies between $(11-15)$~km with mass $(1.4-2)~M_\odot$~\cite{Demorest2010}.
Duncan and Thompson~\cite{duncan/1992} suggest that strong magnetic field is a characteristic
of such highly compact stars. An immense magnetic field of the order of $10^{14}$-$10^{15}$~
Gauss can exist near the surface of the magnetars, strongly magnetized neutron stars
~\cite{thompson/1996,ibrahim/2002}. Later, Chakraborty et al.~\cite{chakraborty/1997} suggest
that, in the core of the neutron star even stronger magnetic field upto $10^{19}$-$10^{20}$~Gauss
can potentially occur. Obviously the reason behind this strong magnetic field is still
under discussion. Though spontaneous ordering of nucleon~\cite{isayev/2004,isayev/2006}
or quark spins~\cite{tatsumi/2000} in the dense interior of a
neutron star can be taken as a cause behind this debating issue.

The extreme high density and tremendous pressure at the core of the
neutron star make it an object of various physical speculation among
the astrophysicists. Due to this high density and
pressure, there is a strong probability of phase transition of neutrons
inside the neutron star to boson, hyperon and strange quark matter.
The prediction of Cameron~\cite{cameron/1959} tells us that hyperon
must be produced inside the neutron star. Due to extreme high density
and weak interaction, some of the nucleon may be converted into hyperon
since this is energetically more favourable. Though interior of the
neutron star may also contain quark matter. The quarks become free
of interaction due to high density and extreme asymptotic momentum transfer
at the core of the neutron star.

An individual nucleon contains quarks to
form a colourless matter known as quark matter. The energy level of
the hyperon at the Fermi-surface becomes higher than its rest mass, due
to the tremendous density, as a result these particles could deconfine
into strange quarks. The strange quarks are the most stable quarks. Though
the quark stars consists of up ($u$), down ($d$) and strange ($s$) quarks
but mostly it contain strange ($s$) quarks. Theoretically under certain
conditions some of the u and d quarks transformed
into s quarks. As we know cold strange matter is the true ground
state of nuclear matter~\cite{Itoh/1970,Bodmer/1971,Witten/1984,Haensel/1986,Alcock1986},
the up ($u$) and down ($d$) quarks once converted into strange
matter, the entire quark matter get converted into strange matter. As a
result the neutron star totally converted into strange quark star~\cite{Pagliara/2013}.
Some recent simulations~\cite{Bauswein/2009,Bauswein/2010} pointed out the merger process
of two strange stars. So their gravitational wave signal detection may enhance
the probability of SQM hypothesis.

Since the quarks are not seen as free particles, the quark confinement
mechanism have been dealt in Quantum Chromodynamics (QCD). A strongly interacting
particle can be defined as a finite region of space, confined with fields~\cite{Chodos/1974}.
Chodos et al.~\cite{Chodos/1974} described the Bag constant ($B_g$) as a finite region with
constant energy density. So, the energy momentum tensor of the star must be effected by this
constant, and thus the geometry of the spacetime also be influenced by the Bag constant.
The MIT bag model states that the reason of quark confinement is due to the universal pressure
$B_g$, known as the MIT bag constant, which is actually the difference between
energy densities of the perturbative and nonperturbative quantum chromodynamical vacuum. For
a stable strange quark matter, the proposed value of the Bag constant should be
within the following range $(55-75)$~MeV/fm$^{3}$~\cite{Farhi1984,Alcock1986}.
However, wide range of the Bag constant are permissible according to CERN-SPS and
RHIC~\cite{burgio/2002}. In a recent work Aziz et al.~\cite{Aziz2019} have shown the
possibility of wide range $(41.58-319.31)$~MeV/fm$^{3}$ for $B_g$ which they have recovered from different observational data for several compact stars. The probable compact strange star candidates are $X-ray$ pulsars, viz. $Her~X1$, millisecond pulsar $SAX~J~1808.43658$, $X$ ray sources, $4U~1820-30$ and $4U~1728-34$ etc.

Lugones and Abra$\tilde{n}$il~\cite{lugones/2017} studied
the structure of strange stars in Randall-Sundrum type~II braneworld model and
investigated the properties of hadronic and strange quark stars with the help
of two typical EOSs, a nonlinear relativistic mean field model for hadronic
matter and the MIT bag model for quark matter.  Abra$\tilde{n}$il and Malheiro
~\cite{arbanil/2016} studied radial stability of anisotropic strange quark
stars, followed the MIT bag EOS, considering vanishing and nonvanishing effect
of anisotropic factor at the surface of a strange star. Abbas et al.~\cite{abbas/2015}
studied the existence of strange stars in $f(T)$ modified gravity with the
help of diagonal tetrad field of static spacetime with charged anisotropic
fluid and MIT bag model. Isayev~\cite{isayev/2015} studied the stability
of magnetized strange quark matter (MSQM) using the MIT bag model with density
dependent bag pressure. Paulucci and Horvath~\cite{paulucci/2014} presented
an analysis of the fragmentation of SQM into strangelets (small lumps of
strange quark matter in which finite effects become important)~\cite{madsen/1999,madsen/2002}
in high temperature astrophysical events, considering quark matter within
the MIT bag model framework in color-flavor-locked (CFL) state~\cite{alford/1999,rapp/2000,lugones/2002}.
Deb et al.~\cite{deb/2017} studied ultra dense strange (quark) stars employing
MIT bag model, assuming density profile as provided by Mak and
Harko~\cite{mak/2002}. They featured an interesting result that anisotropy
of highly compact strange stars increases with radial coordinate and attains its
maximum value at the surface. This result seems to be an inherent property
of the singularity free anisotropic strange stars. Rahaman et al.~\cite{Rahaman/2012}
studied anisotropic charged strange star in GR using MIT bag EOS, where $B_g$
value becomes relatively higher probably to balance the excess repulsion, caused by
electric field.

Recently, anisotropic strange star in modified $f(R,T)$ gravity has
been investigated by Deb et al.~\cite{Deb/2018b} when
the matter geometry coupling is simplest and minimum. They also studied isotropic
case~\cite{Deb/2018a} in $f(R,T)$ gravity where they provided a stable stellar model
to study strange stars. In our previous work~\cite{Biswas/2018}, we have introduced a new model
for highly compact anisotropic strange star using the metric potentials, given by
Krori-Barua~\cite{Krori/1975} under $f(R,T)$ gravity. We have calculated and explained an interesting
result that value of the bag constant $B_g$ reduces in $f(R,T)$ gravity due to the coupling
effect between matter and geometry. Besides, there are so many more literatures
~\cite{Brilenkov/2013,Maharaj/2014,Panda/2015,Bhar/2015} available on strange stars based on MIT bag model EOS.

Depending on the inhomogeneity of matter distribution and its evolution, which
also results anisotropic features, the fabric of spacetime of compact object
changes. During the investigation of stellar properties we must have to
keep in our mind that, for smaller radial size with higher density, anisotropic
pressure plays an important role. The term anisotropic pressure signifies that
the radial component of pressure $(p_r)$ differ from that of tangential component
of pressure $(p_t)$. Actually Ruderman~\cite{ruderman/1972} has investigated
the stellar models and argued that the nuclear matter
may have anisotropic features at very high density $(> 10^{15}$ gm/cc), where
the nuclear interaction must be treated relativistically. After this
remarkable concept Bower and Liang~\cite{bower/1974}, Bayin~\cite{bayin/1982},
Maharaj and Maartens~\cite{maharaj/1989} emphasised on the
importance of locally anisotropic EOS for relativistic fluid
spheres. Actually anisotropy affects the physical properties of a stellar object
such as the radial pressure, total radius, total mass, energy density, surface redshift
and the frequency of oscillation of the fundamental mode.

According to Kippenhahn and Weigert~\cite{kippenhahn/1990}, the anisotropy may
arises due to  presence of a type $3A$ superfluid or existence of solid stellar
core. Sawyer~\cite{sawyer/1972} suggested that anisotropy in pressure may generate
due to pion condensation. Another reason for anisotropy may be taken as different
kind of phase transitions, according to Sokolov~\cite{sokolov/1980}.
According to Weber~\cite{weber/1999} the immense magnetic field of neutron star
may also produce anisotropy in pressure inside a strange star. Also the strong
electric field produces the effect of anisotropy, as suggested by Usov~\cite{usov/2004}.
The interior gravitational fields of anisotropic fluid spheres can be described
properly by exact solutions of the EFE as shown in following references
~\cite{krori/1984,ivanov/2002,schunck/2003,mak/2003,varela/2010,rahaman/2010,rahaman/2011,maurya/2012,kalam/2012,rahaman/2012,maurya/2015,malaver/2015,pant/2015,maurya/2016,shee/2016,shee/2017}.

To study the anisotropic strange star, we employ the TK metric potentials
which provide a class of singularity free solution. Tolman in 1939~\cite{tolman/1939}
independently gave analytical solutions of EFE by choosing eight
different types of metric potentials. Also Kuchowicz in 1968~\cite{kuchowicz/1968} independently
gave this singularity-free metric potential to describe the stellar configuration.

Recently, Jasim et al.~\cite{jasim/2018} presented a study on anisotropic strange star under the framework of Einstein's General
Relativity by using the TK metric potentials. It is worthy to mention that their result provides
theoretical support for massive strange stars for specific values of the model parameters.
However, though flavour of MIT bag EOS is hidden in the value of $B_g$, the authors have considered the value for it arbitrarily.
In connection to this aspect one may finds in literature~\cite{Farhi1984,Alcock1986} that for stable SQM with the proposed range is
$(55-75)$~MeV/fm$^3$.

Being motivated by the above background, especially from the work of Jasim et al.~\cite{jasim/2018},
here we have tried to introduce TK type metric potentials which provide us a singularity-free
stellar model for a wide range of MIT bag values instead of a specific one. The scheme of the study
is as follows: In Section 2 we have provided basic mathematical formulation of Einstein's spacetime.
Section 3 contains solutions to the field equations. To represent the model properly
we have found the numerical values of different model parameters, which are provided in
Section 4. The physical features of our model are represented in Section 5 through
various subsections, namely, density and pressure, Equation of State, TOV equation,
energy conditions, Herrera's Causality and Cracking concept, adiabatic index and
Harrison-Zel$'$dovich-Novikov static stability criteria. Section 6 contains mass-radius
relation through which we can represent surface redshift of an anisotropic strange
star. Lastly, in Section 7 we have drawn some concluding remarks on different
aspects of the present model.

\section{Basic Mathematical Formulation of Einstein's Spacetime}
We consider static spherically symmetric spacetime metric through the line element
\begin{equation}
ds^2=e^{\nu(r)}dt^2-e^{\lambda(r)}dr^2-r^2(d\theta^2+\sin^2\theta d\phi^2),\label{eq1}
\end{equation}
where $\lambda(r)$ and $\nu(r)$ are metric function with TK type~\cite{tolman/1939,kuchowicz/1968},
chosen as $\lambda(r)=\ln(1+ar^2+br^4)$ and $\nu(r)=Br^2+2\ln C$. Here $a$, $b$, $B$ and $C$ are arbitrary
constants which can be evaluated on the basis of several physical requirements. These potentials are well behaved,
satisfy the criteria for physical acceptance and free from the central singularity.

The energy-momentum tensor for anisotropic matter distribution,
compatible with the spherically symmetric spacetime with signature (+, -, -, -), is given by
\begin{eqnarray}
{T_{{\nu}}}^{\mu}= \left( \rho+p_{r} \right) u^{{\mu}}
{u}_{\nu} + p_{r}{g_{\nu}}^{\mu}+ \left( p_{t}-p_{r} \right) {\eta}^{\mu}{\eta}_{\nu}, \label{eq2}
\end{eqnarray}
with $u^{i}u_{j}=-{\eta}^{i}{\eta}_{j}=1$ and $u^{i}{\eta}_{j}=0$. Here the
vector $u_{i}$ is the $4$-velocity and ${\eta}^{i}$ is the spacelike
vector which is orthogonal to $u_{i}$. Here $\rho$, $p_r$ and $p_t$ are the matter
density, radial pressure and tangential pressure respectively, of the
fluid which is in orthogonal direction to $p_r$. Using Eqs.~(\ref{eq1})
and~(\ref{eq2}) one can obtain Einstein's field equations, assuming $G=c=1$ (i.e. Geometrized unit), as follows
\begin{eqnarray}
&\qquad\hspace{-2.8cm}8\pi\rho=-e^{-\lambda}\left[\frac{1}{r^2}-\frac{\lambda'}{r}\right]+\frac{1}{r^2}, \label{eq3}\\
&\qquad\hspace{-3.0cm}8\pi p_r=e^{-\lambda}\left[\frac{\nu'}{r}+\frac{1}{r^2}\right]-\frac{1}{r^2},  \label{eq4}\\
&\qquad\hspace{-0.5cm}8\pi p_t=\frac{e^{-\lambda}}{4}\left[2\nu''+{\nu'}^2-\nu'\lambda'+\frac{2}{r}(\nu'-\lambda')\right], \label{eq5}
\end{eqnarray}
where `$\prime$'  denotes the derivatives of the respective parameters with
respect to the radial parameter $r$.

The simplest phenomenological MIT bag model EOS \cite{chodos/1974,Farhi1984,Alcock1986}
governs the SQM distribution inside the strange star. Considering that quarks are non-interacting,
massless and including all the corrections of energy and pressure, the quark
pressure of SQM can be defined as
\begin{equation}
p_r(r)=\sum_{f=u,d,s} p^{f}-B_{g}, \label{eq6}
\end{equation}
where $p^{f}$ is the individual pressure of all three flavors of quarks
and $B_{g}$ is also known as vacuum energy density
which can be treated as a constant quantity though recent research support
its wide range of values.

The relation between the individual quark pressure $p^{f}$ and energy density
of individual quark flavor is given by $p^{f}=\frac{1}{3}\rho^{f}$.
Therefore deconfined quarks inside the bag have the total energy $\rho$ as follows
\begin{equation}
\rho=\sum_{f=u,d,s} \rho^{f}+B_g, \label{eq7}
\end{equation}
using Eqs.~(\ref{eq6}) and~(\ref{eq7}) we have the MIT bag EOS as
\begin{equation}
p_r(r)=\alpha\left[\rho(r)-4B_g\right], \label{eq8}
\end{equation}
where $\alpha$ is a constant. Jasim et al.~\cite{jasim/2018} considered $\alpha=0.28$
for the massive strange quarks having mass 250~MeV \cite{Stergioulas/2003},
however for massless strange quarks, it has the value $\frac{1}{3}$ which we have
considered in the present investigation for strange star. But we scientifically have studied it
by finding out the values of $B_g$ in terms of the mass-radius of different strange star candidates
where depending on the values of these quantities it can be seen that $B_g$ will change in a characteristic way.

Without involving any quantum mechanical particle aspect, this simplest EOS
is very useful to study the equilibrium configuration of a
compact star, in the framework of General Theory of Relativity, made up of
only up, down and strange quarks. As we know the
radial pressure must vanish at the surface so from Eq.~(\ref{eq8}) we have
\begin{equation}
\rho_R=4B_g, \label{eq9}
\end{equation}
where $R$ is the radius of the strange star and $\rho_R$ is the surface
density of the compact stellar model.

\section{Solutions of the Field Equations}
Now using Eq.~(\ref{eq8}) along with the metric functions $\nu$ and $\lambda$, we can solve
Eqs.~(\ref{eq3})-(\ref{eq5}) and eventually we have the following solutions
\begin{eqnarray}
&\qquad\hspace{-1.1cm}\rho=\frac{3}{16}\frac{Bbr^4+Bar^2+2br^2+B+a}{\pi(br^4+ar^2+1)^2}+B_g, \label{eq10}\\
&\qquad\hspace{-1.0cm}p_r=\frac{1}{16}\frac{Bbr^4+Bar^2+2br^2+B+a}{\pi(br^4+ar^2+1)^2}-B_g,  \label{eq11}\\
&\qquad\hspace{-.2cm}p_t=\frac{1}{8}\frac{B^2br^6+B^2ar^4+B^2r^2+Bar^2-2br^2+2B-a}{\pi(br^4+ar^2+1)^2}. \label{eq12}
\end{eqnarray}

The anisotropy ($\Delta$) of our system reads as
\begin{eqnarray}
&\qquad\hspace{-3.1cm}\Delta \equiv (p_t-p_r)=\frac{(Br^2+\frac{3}{2})(Bbr^4+(Ba-2b)r^2+B-a)}{8\pi(br^4+ar^2+1)^2}+B_g. \label{eq13}
\end{eqnarray}

The metric potentials can be expressed as
\begin{eqnarray}
&\qquad\hspace{+.1cm}e^{\lambda(r)}=br^4+ar^2+1, \label{eq14}\\
&\qquad\hspace{-0.1cm}e^{\nu(r)}=e^{Br^2+2\ln(C)}. \label{eq15}
\end{eqnarray}

Variation of the metric potentials $e^{\nu}$ and $e^{\lambda}$ are shown in
Fig.~\ref{pot.} (upper and lower panel respectively) for various strange stars.
From these graphical representations, it is very clear that $e^{\nu(r)}\vert_{r=0}$
is non zero positive value and  $e^{\lambda(r)}\vert_{r=0}=1$ which are the
necessary conditions for the solution to be free from physical as well as
geometric singularities. Both the metric potential are minimum at the centre,
then increase nonlinearly and become maximum at the surface.

\section{Bounds on the model parameters}

\subsection{Interior spacetime}
The central density can be obtained from Eq.~(\ref{eq10}) as follows
\begin{eqnarray}
&\qquad\hspace{-0.2cm}\rho_c=\rho(r=0)=\frac{3}{16\pi}(B+a)+B_g. \label{eq16}
\end{eqnarray}

Anisotropic condition states that, at the center ($r=0$) anisotropy is zero, i.e., $p_r=p_t$,
\begin{eqnarray}
&\qquad\hspace{-0.2cm}p_r(r=0)=p_t(r=0).\label{eq17}
\end{eqnarray}

\subsection{Exterior spacetime}
To find out the values of constants, we will match our interior spacetime with
the exterior Schwarzschild metric~\cite{schwarzschild/1916} which is given by
\begin{eqnarray}
&\qquad\hspace{-.5cm}ds^2=\left(1-\frac{2M}{r}\right)dt^2-\left(1-\frac{2M}{r}\right)^{-1}dr^2-r^2(d\theta^2+\sin^2\theta d\phi^2),    \label{eq18}
\end{eqnarray}
where $M$ is the total mass of the strange star.

At the boundary, i.e., $r=R$ where $R$ is the radius of the corresponding star,
metric coefficients $g_{tt}$, $g_{rr}$ and $\frac{\partial g_{tt}}{\partial r}$ are
continuous between the interior and exterior region which leads to the following conditions
\begin{eqnarray}
&&g_{tt}:1-\frac{2M}{R}=e^{BR^2+2\ln C},\label{eq19} \\
&&g_{rr}:(1-\frac{2M}{R})^{-1}=(1+aR^2+bR^4), \label{eq20} \\
&&\frac{\partial g_{tt}}{\partial r}:\frac{M}{R^2} = BRe^{BR^2+2\ln C}. \label{eq21}
\end{eqnarray}

Solving Eqs. (\ref{eq19})-(\ref{eq21}), we can obtain the values
of the constants $a$, $b$, $B$, $C$ in terms of $M$ and $R$ as follows
\begin{eqnarray}
a&=&-\frac{1}{2}\frac{M\left(8M^2-9MR+4R^2\right)}{R^2(-R+2M)(M^2-MR+R^2)} ,\label{eq22} \\
b&=&\frac{4M^3-5M^2R}{4M^3R^4-6M^2R^5+6MR^6-2R^7},\label{eq23} \\
B&=&\frac{M}{R^2(R-2M)},\label{eq24}  \\
C&=&e^{\frac{(-R+2M)\ln\left({\frac{R-2M}{R}}\right)+M}{-2R+4M}}.\label{eq25}
\end{eqnarray}

At the surface (i.e., $r=R$) radial pressure vanishes, that gives
\begin{eqnarray}
&\qquad\hspace{-.7cm}p_{r}(r=R)=\left[\frac{(BbR^4+BaR^2+2bR^2+B+a)}{16\pi(R^4b+R^2a+1)^2}\right]-B_g=0. \label{eq26}
\end{eqnarray}

Inserting the expressions for $a$, $b$, $B$ in Eq. (\ref{eq26}), $B_g$ becomes
\begin{equation}
B_g=\frac{3M(2R-3M)}{32\pi R^2(M^2-MR+R^2)}.  \label{eq27}
\end{equation}

Finally we have obtained all constants in terms of mass ($M$) and radius ($R$)
of the strange star. Using observed values of these quantities
we have calculated numerical values of different constants and
physical parameters for various candidates of strange stars, shown in Tables 1 and 2.

\begin{figure}[!htbp]
\centering
\includegraphics[width=6cm]{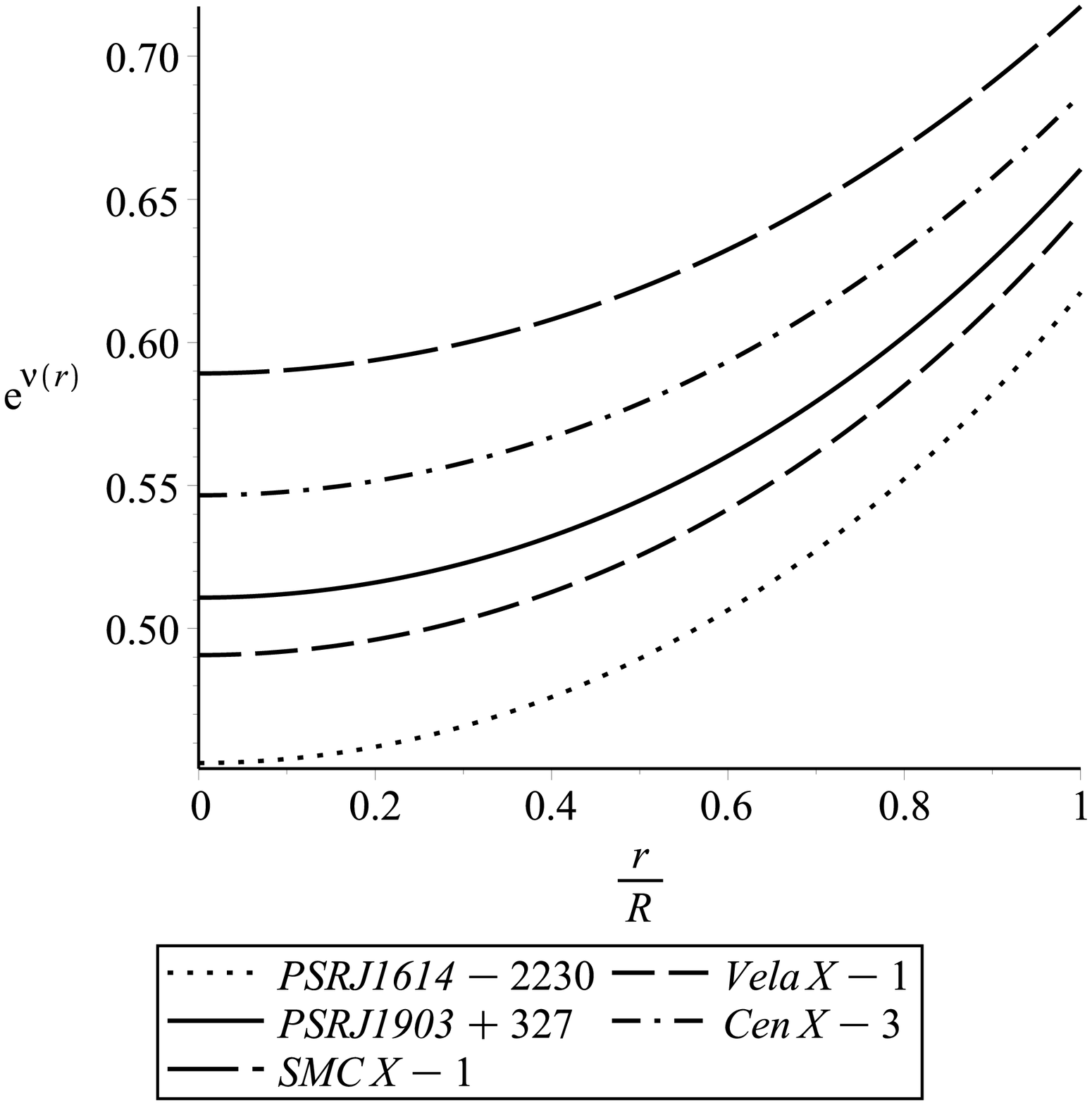}
\includegraphics[width=6cm]{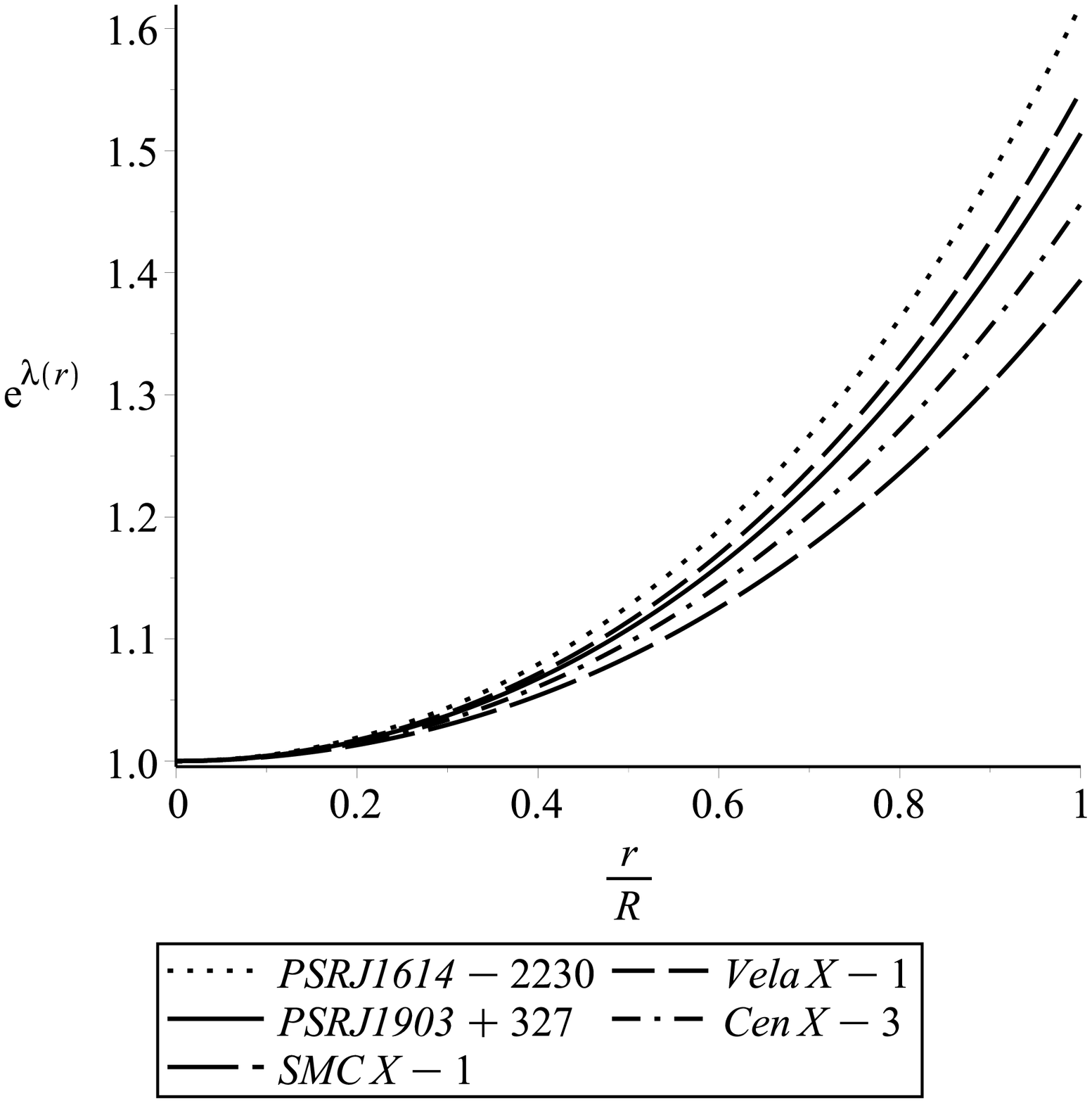}
\caption{Variation of $e^{\nu}$ (left panel), $e^{\lambda}$ (right panel) w.r.t.
the fractional radial coordinate $r/R$ for different strange star candidates.}\label{pot.}
\end{figure}

\section{Physical features of the proposed model}

\subsection{Density and pressure}
\begin{figure}[!htbp]
\centering
\includegraphics[width=6cm]{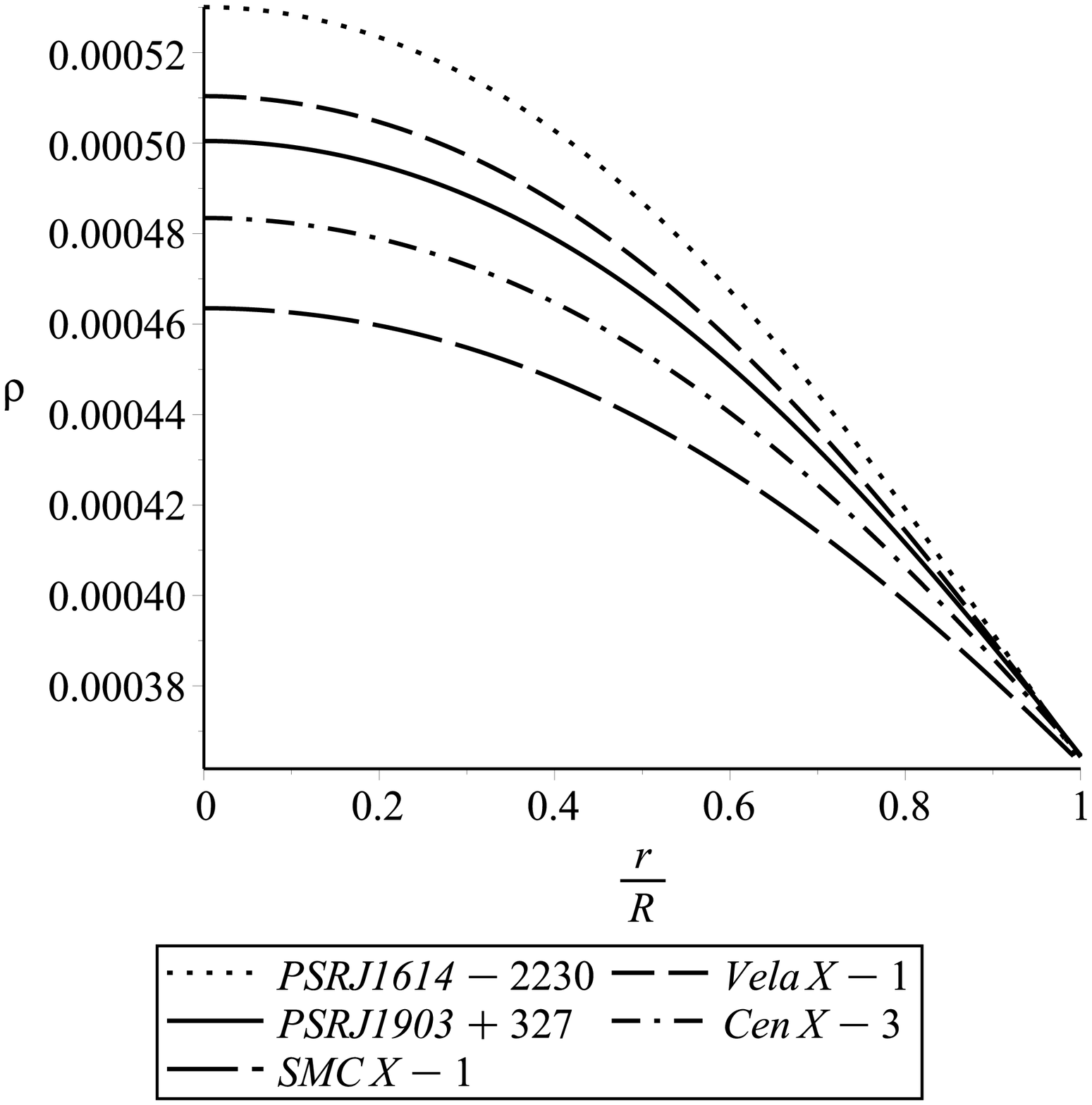}
\includegraphics[width=6cm]{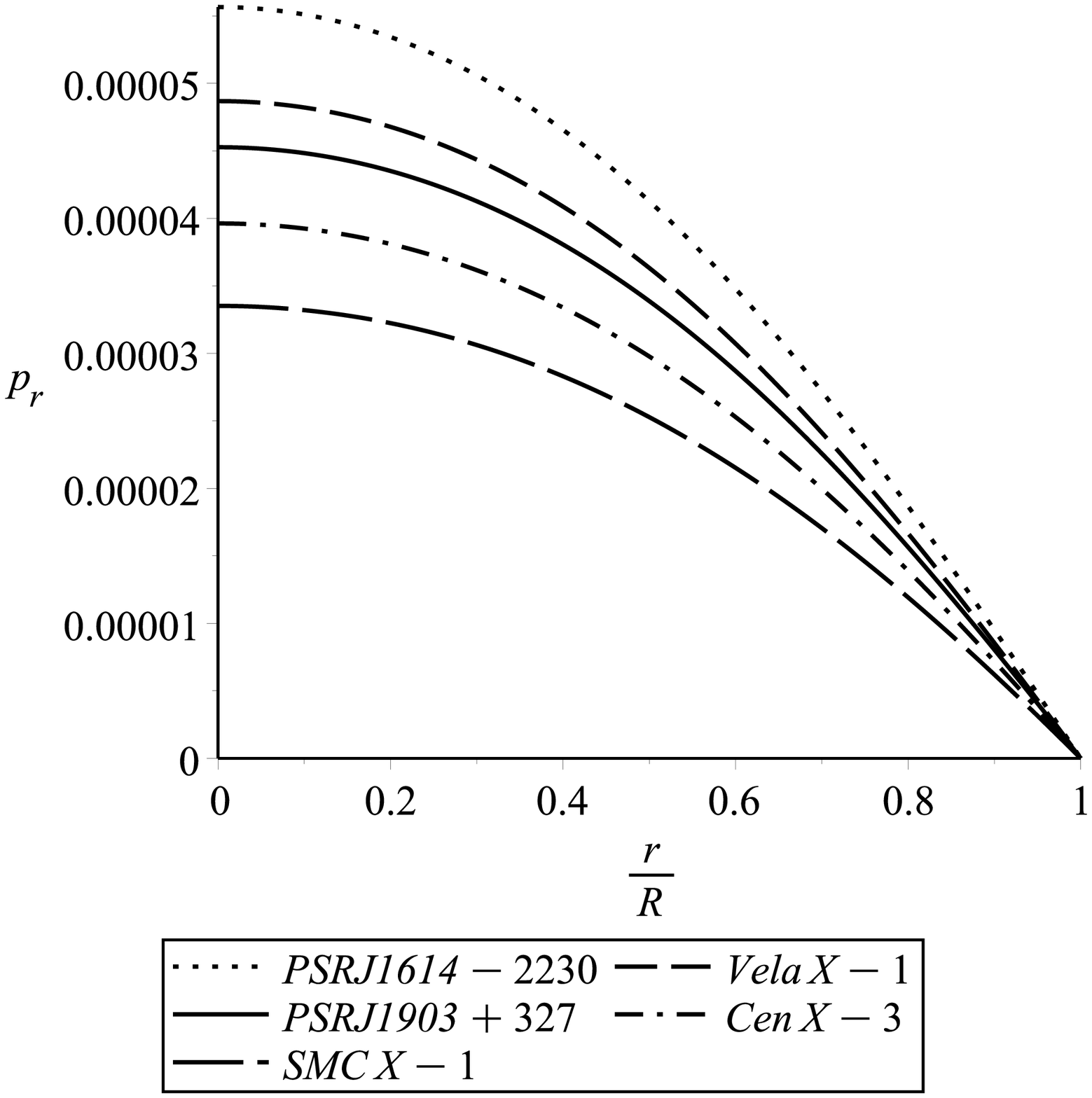}
\includegraphics[width=6cm]{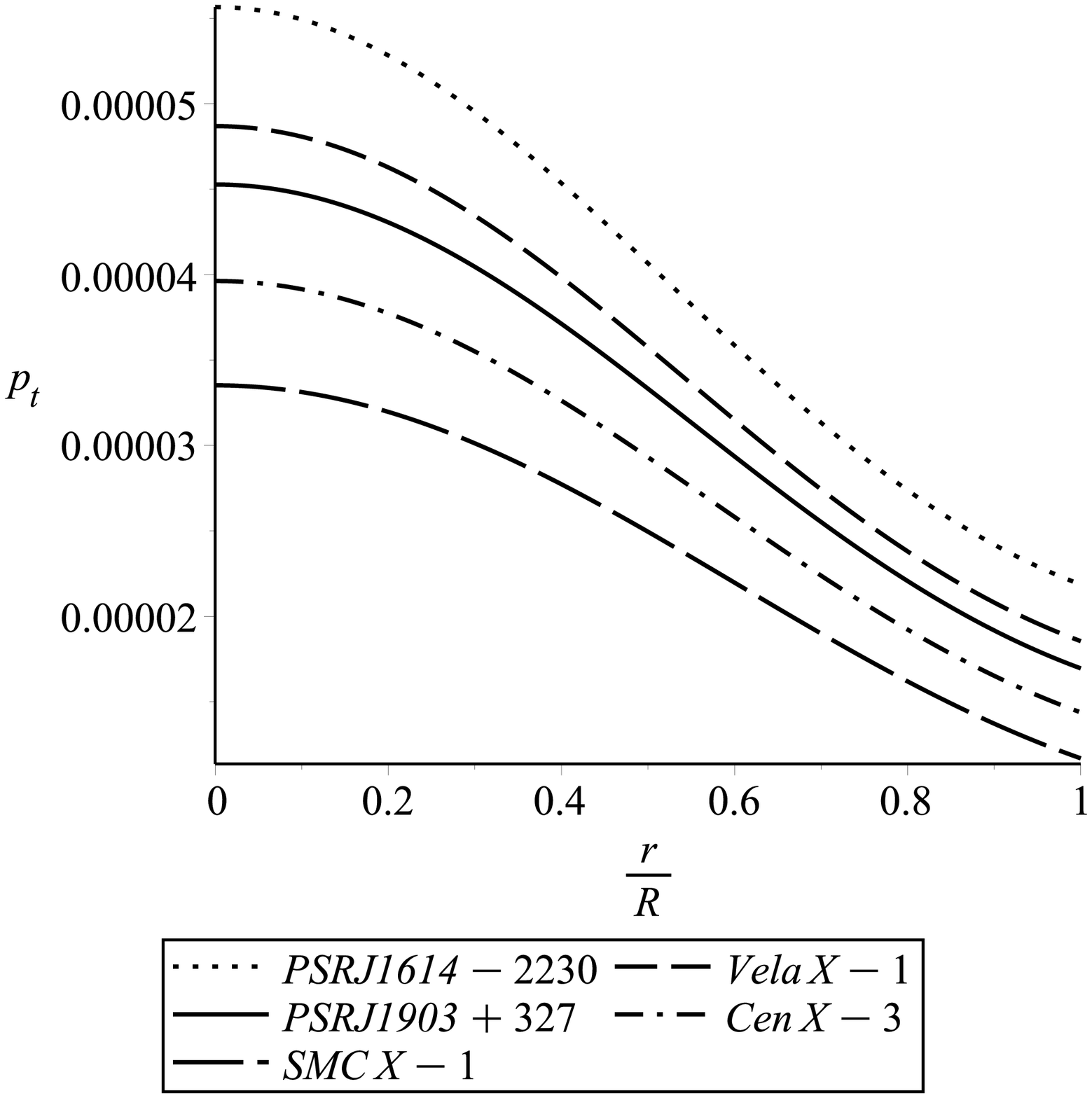}
\includegraphics[width=6cm]{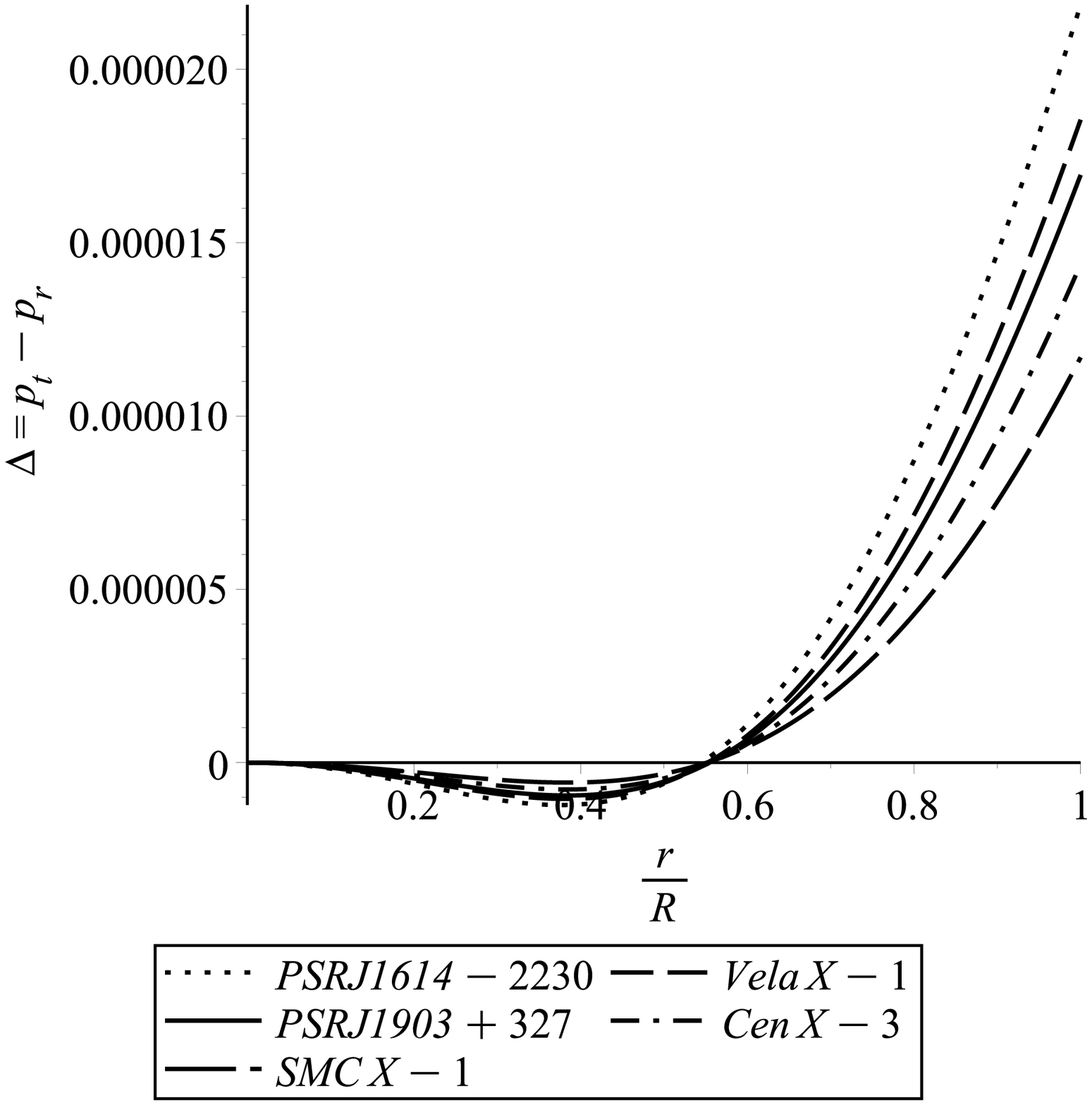}
\caption{Variation of density (upper left), radial pressure (upper right), tangential pressure (lower left) and anisotropic stress (lower right)
w.r.t. the fractional radial coordinate $r/R$ for different strange star candidates.}\label{pres.}
\end{figure}

Using density profile as expressed in Eq.~(\ref{eq10}) along with the numerical values of different
constants, we are now able to plot the variation of $\rho(r)$ with fractional radial coordinate $r/R$ for the different
strange star candidates. From Fig.~\ref{pres.}, it is clear that
density is finite at the center of the strange star and decreases monotonically with
the increase of radius of the stellar body. If we look at the numerical values
of central density and surface density of various candidates of strange
stars in Table 1, then table transparently shows the much higher value of central density
w.r.t. the density at the surface of the stars, which is the characteristic of ultra dense
strange quark stars~\cite{ruderman/1972,Glendenning/1997,Herzog/2011}.

The radial pressure $p_r$ and tangential pressure $p_t$ are also expressed
graphically using the numerical values of different constants.
Their finite values at the center certify a singularity free model of strange
star. From Fig.~\ref{pres.} we can see that radial pressure vanishes
at the boundary of stars but tangential pressure dose not vanishes sharply at
the surface of the star which indicate the spheroidal nature of the
strange stars~\cite{quevedo/1989,chifu/2012,shee/2017}.

We have defined the anisotropy ($\Delta$) of our model in Eq.~(\ref{eq13}).
According to Hossein et al.~\cite{Hossein/2012}, anisotropy will be directed outward when
$\Delta>0$  i.e., $p_t>p_r$, and inward when $\Delta<0$ i.e., $p_t<p_r$. From
Fig.~\ref{pres.} it is clear that anisotropy vanishes at the center of the star
after that it remain negative upto certain distance. According to Hossain
et al.~\cite{Hossein/2012}, negative anisotropy allows the construction of massive
stellar structure. It is very much prominent from the graphical representation
that anisotropy will become positive after overcoming the negative value.
According to Gokhroo and Mehra~\cite{gokhroo/1994} positive anisotropy helps
to construct the more compact object. So our model acquire a stable
configuration in the frame work of anisotropy. Also from the graphical
representation we can see that the anisotropy increases and attain its maximum
value at the surface of the star which is an inherent property of an ultra
dense compact star~\cite{deb/2017}.

\subsection{Equation of State (EOS)}
The radial ($\omega_r$) and tangential ($\omega_t$) EOS parameters for our
system can be expressed as
\begin{eqnarray}
&\qquad\hspace{-2.2cm}\omega_r(r)=\frac{p_r}{\rho}=\frac{\alpha-16\pi B_g(br^4+ar^2+1)^2}{3\alpha+16\pi B_g(br^4+ar^2+1)^2}, \label{eq28} \\
&\qquad\hspace{-.5cm}\omega_t(r)=\frac{p_t}{\rho}=2\frac{B^2r^2(br^4+ar^2+1)-Bar^2-2br^2+2B-a}{3\alpha+16\pi B_g(br^4+ar^2+1)},\label{eq29}
\end{eqnarray}
where, $\alpha=(Bbr^4+Bar^2+2br^2+B+a)$.

\begin{figure}[!htp]
\centering
\includegraphics[width=6cm]{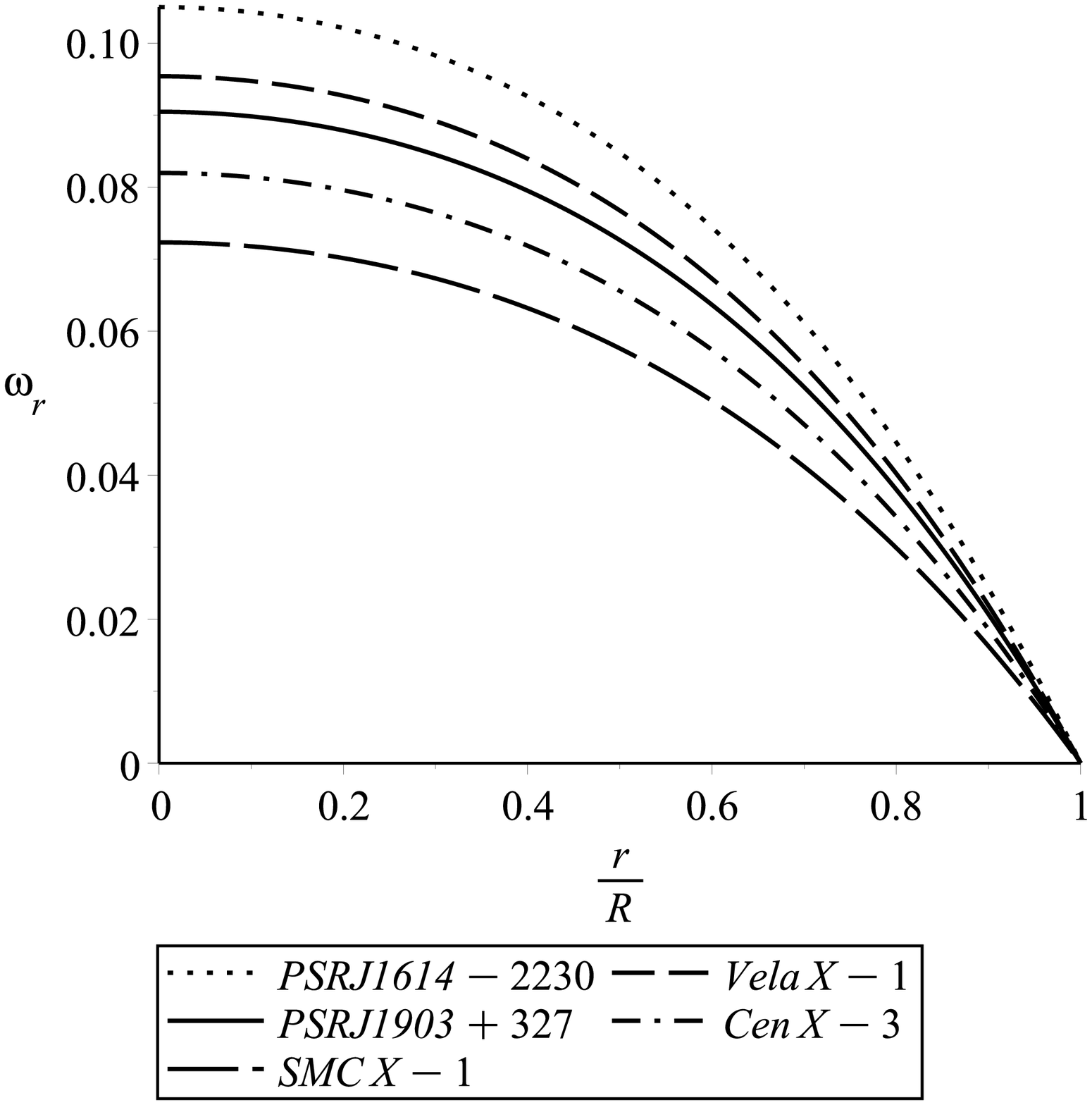}
\includegraphics[width=6cm]{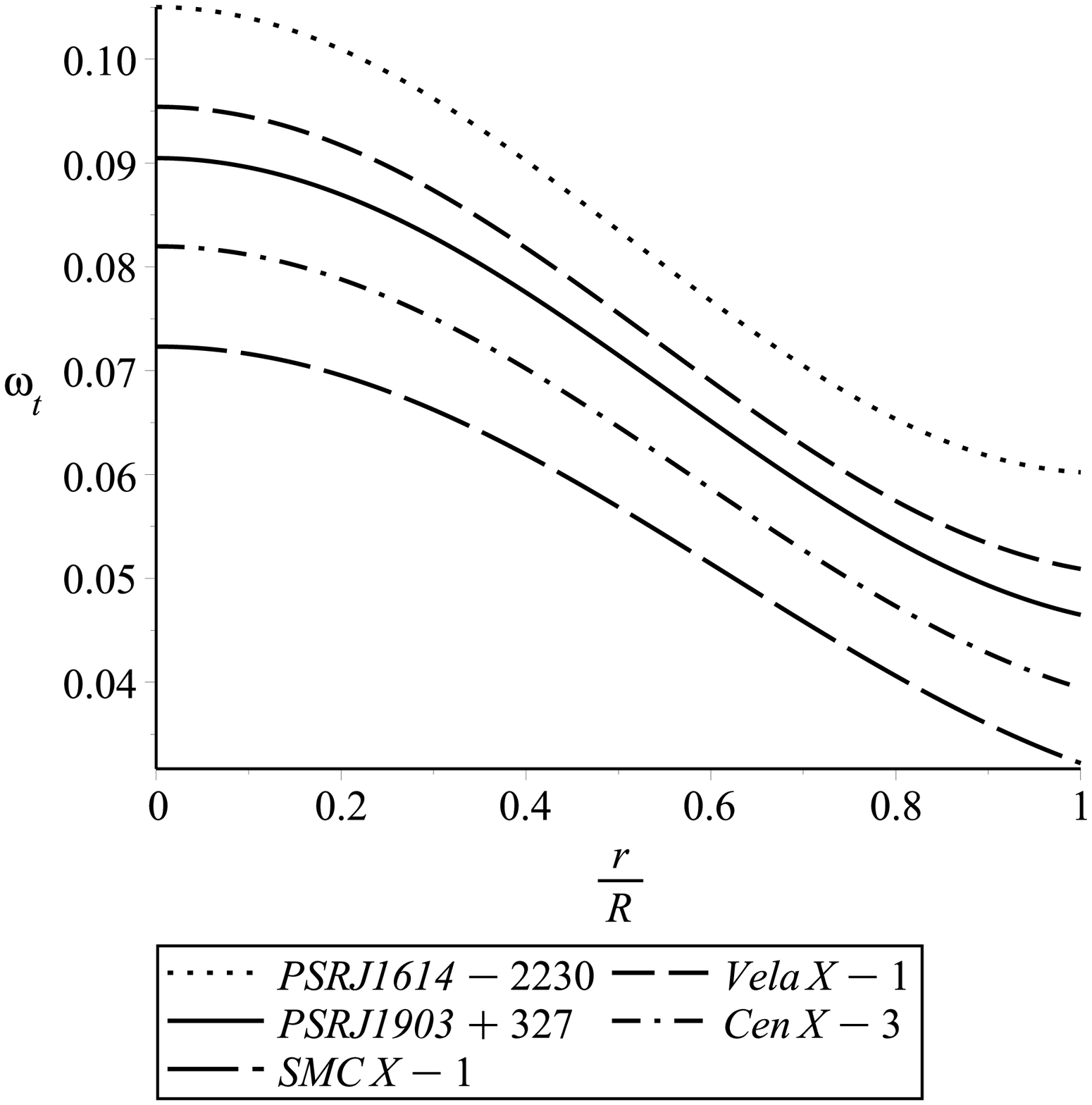}
\caption{Variation of equation of state $w_r$ (left panel) and $w_t$ (right panel)
w.r.t. the fractional radial coordinate $r/R$ for different strange star candidates.}\label{eos.}
\end{figure}

From Fig.~\ref{eos.} it is clear that $0<\omega_i<\frac{1}{3}$ and decreases
towards the surface of the respective stars. The inequality
condition $0<\omega_i<\frac{1}{3}$ indicate the non-exotic nature~\cite{shee/2016} of the underlying
fluid distribution. Here we have considered the simplest barotropic EOS as $p_i=\omega_i\rho$,
where $\omega_i$ are the EOS parameters along the radial and tangential directions.
Here we have considered spatially homogeneous cosmic fluid. Though space and
time dependence of $\omega$ are also possible which are in the following literature
survey~\cite{chervon/2000,zhuravlev/2001,peebles/2003,usmani/2008}.

\subsection{The generalised Tolman-Oppenheimer-Volkoff equation}
To study the stability of a strange star under different forces we have to put our model under generalised Tolman-Oppenheimer-Volkoff (TOV)
equation as proposed by Ponce de Le\'{o}n~\cite{Leon/1987}, which can be represented as
\begin{equation}
-\frac{M_G(\rho+p_r)}{r^{2}}e^{\frac{\lambda-\nu}{2}}-\frac{dp_r}{dr}+\frac{2}{r}(p_t-p_r)=0,\label{eq30}
\end{equation}
where $M_G=M_G(r)$ is the effective gravitational mass inside a sphere of radius $r$. $M_G(r)$ can be derived from the modified
Tolman-Whittaker formula~\cite{devitt/1989} as
\begin{equation}
M_G(r)=\frac{1}{2}r^{2}e^{\frac{\nu-\lambda}{2}}\nu'.\label{eq31}
\end{equation}

Substituting this value in Eq.~(\ref{eq30}) we get the following form of TOV equation
\begin{eqnarray}
&\qquad\hspace{-.1cm}-\frac{(\rho+p_r)\nu'}{2}-\frac{dp_r}{dr}+\frac{2}{r}(p_t-p_r)=0. \label{eq32}
\end{eqnarray}

The above equation explains the equilibrium conditions of the considered
fluid sphere under the combined effect of gravitational, hydrostatic and anisotropic forces:
\begin{equation}
F_g+F_h+F_a=0. \label{eq33}
\end{equation}

Here
\begin{eqnarray}
&\qquad\hspace{-1.3cm}F_g=-\frac{Br(Bbr^4+Bar^2+2br^2+B+a)}{4\pi(br^4+ar^2+1)^2},  \label{eq34} \\
&\qquad\hspace{-.5cm}F_h=\frac{(Bbr^4+Bar^2+2br^2+B+a)(2br^3+ar)}{4\pi(br^4+ar^2+1)^3}-\frac{(2Bbr^3+Bar+2br)}{8\pi(br^4+ar^2+1)^2},  \label{eq35}\\
&\qquad\hspace{+0.2cm}F_a=\frac{[Bbr^4+(Ba-2b)r^2+B-a](Br^2+3/2)}{4\pi r(br^4+ar^2+1)^2}+\frac{2B_g}{r}. \label{eq36}
\end{eqnarray}

\begin{figure}
\centering
\includegraphics[width=4.4cm]{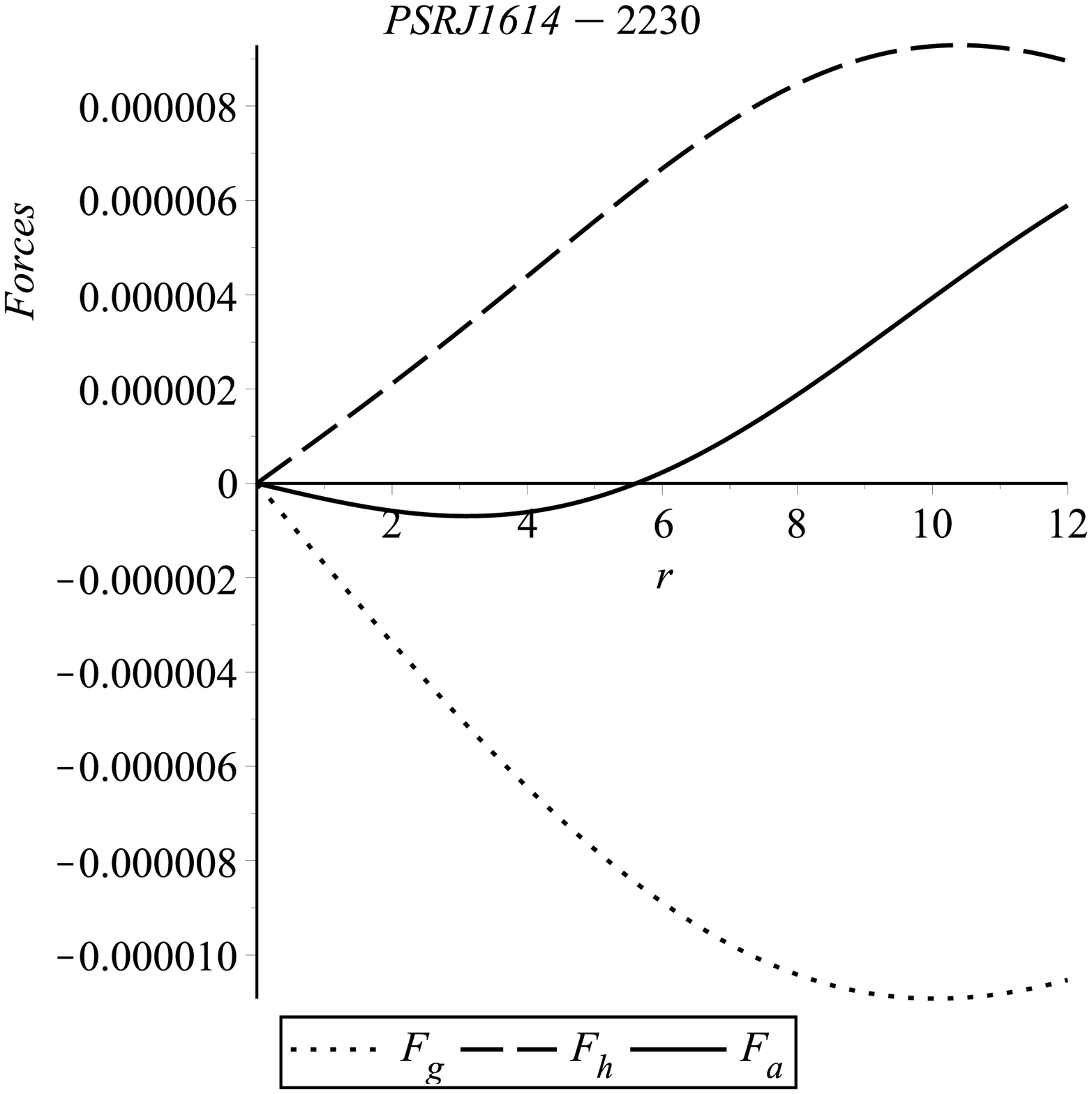}
\includegraphics[width=4.4cm]{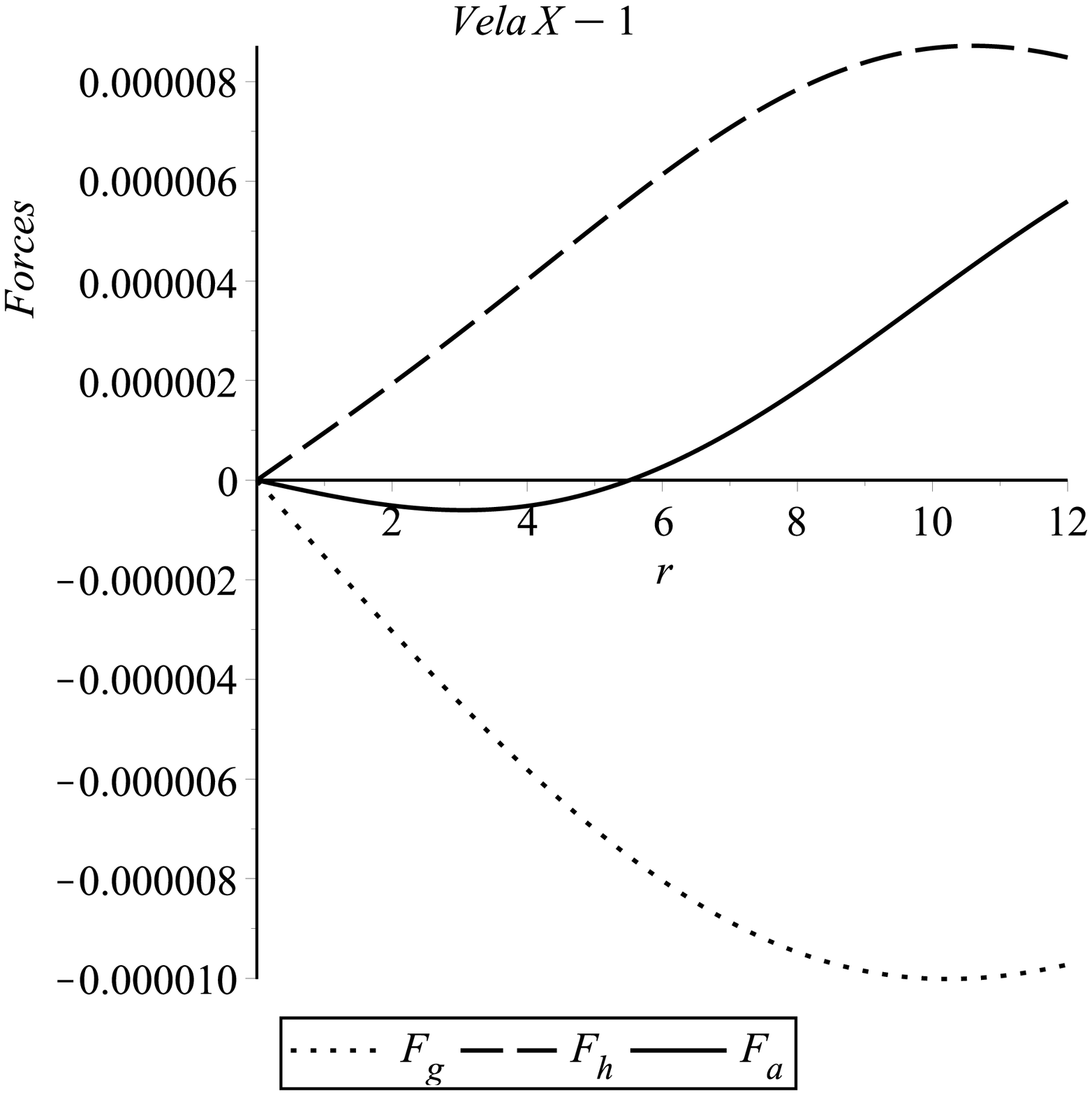}
\includegraphics[width=4.4cm]{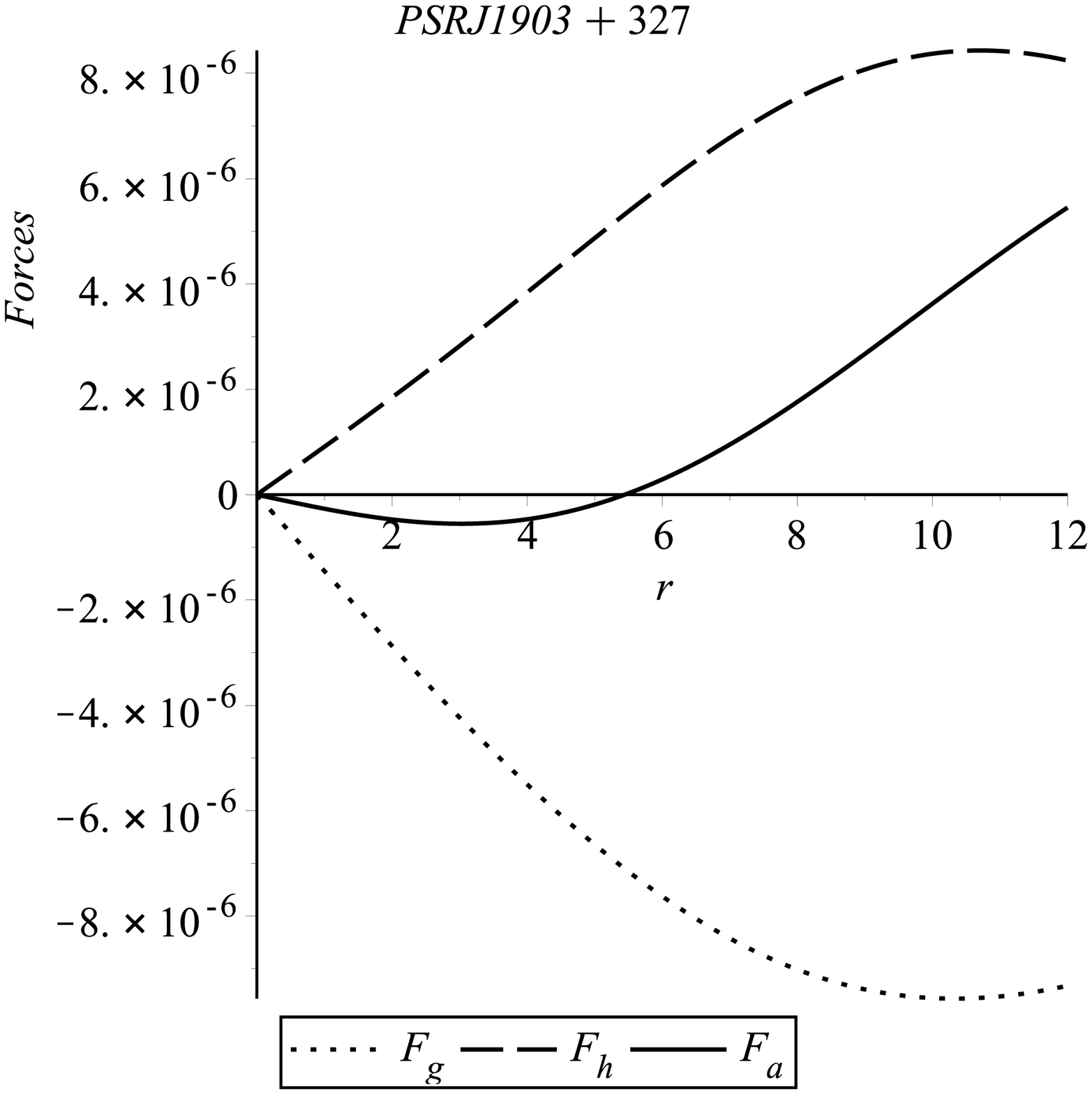}
\includegraphics[width=4.5cm]{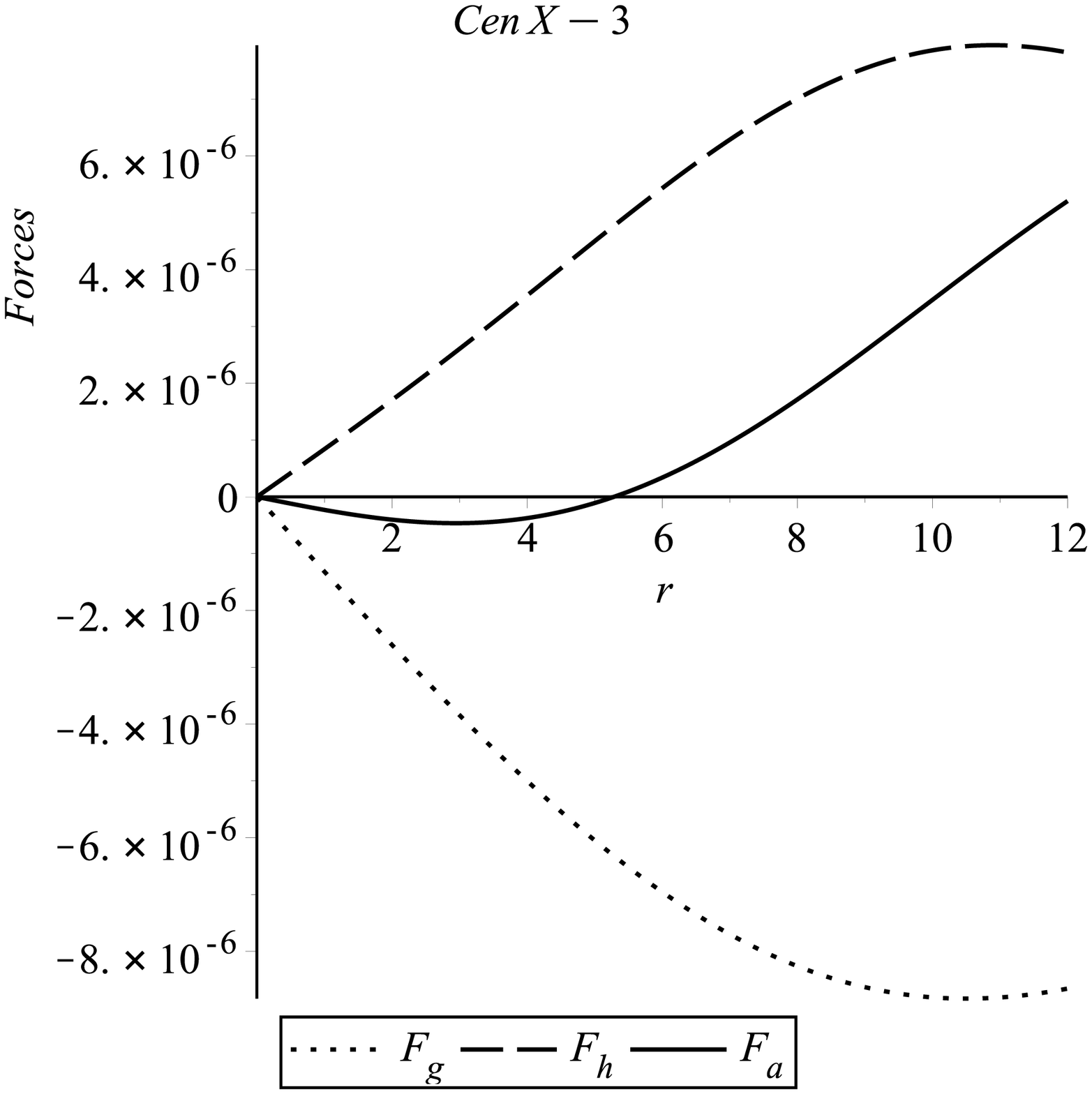}
\includegraphics[width=4.5cm]{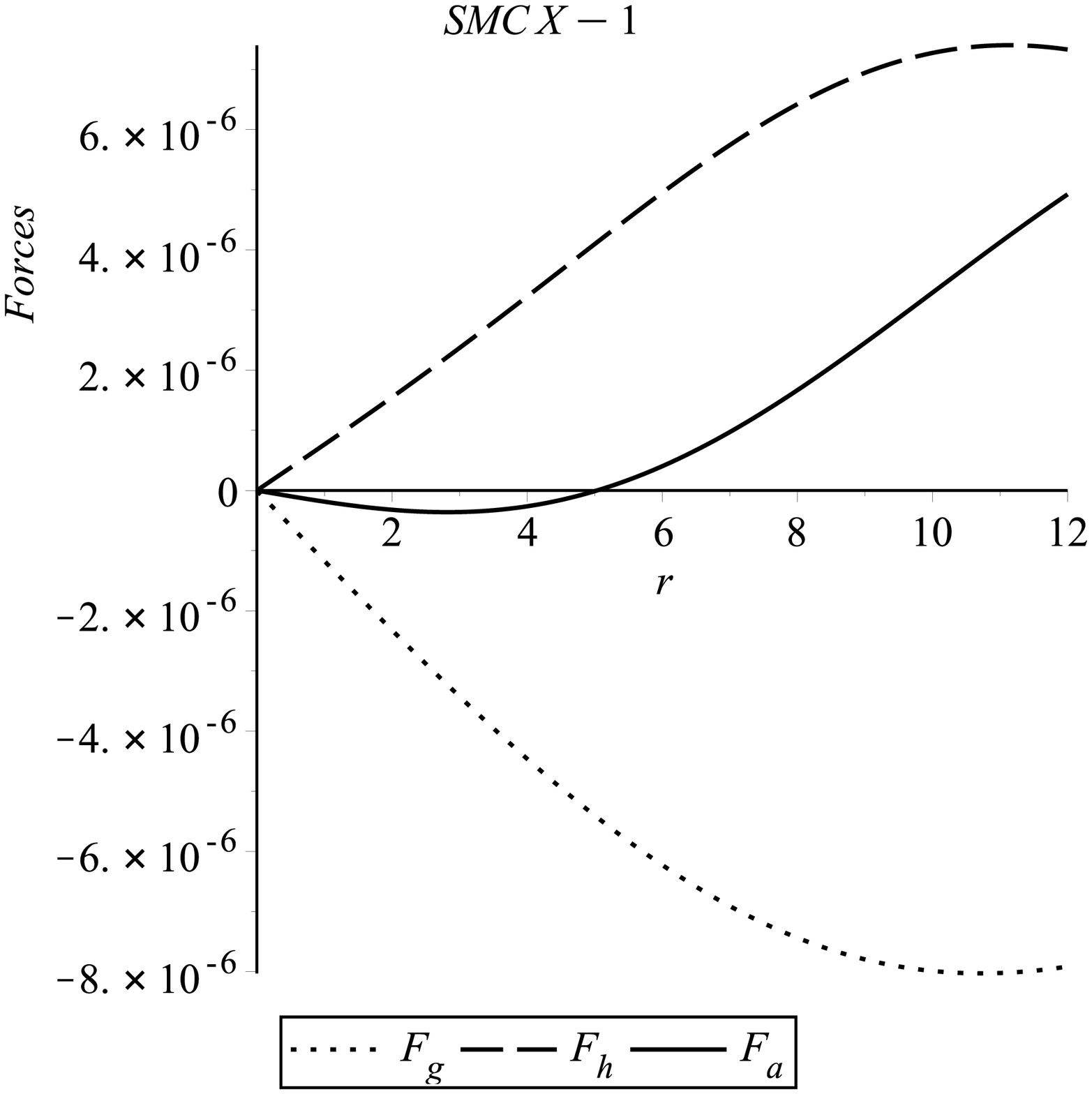}
\caption{Variation of different forces w.r.t. the radial coordinate $r$ for
different strange star candidates.}\label{tov.}
\end{figure}

The variations of different forces $F_g$, $F_h$ and $F_a$ are shown in Fig.~\ref{tov.}.
The profile tells us that the combined effect of hydrostatic force
and anisotropic force is counter balanced by the gravitational force, as a
result our considered ultra dense compact stars are in stable equilibrium situation.

\subsection{Energy Conditions}
In GR, the energy-momentum tensor $T_{\nu}^{\mu}$ is described as mass, momentum and stress
distribution caused by matter as well as any non-gravitational fields. However, the EFE is
flexible enough about the acceptance of non-gravitational fields and states of matter
in a spacetime model. Hence, the energy conditions represent and describe the common
properties for various states of matter and all well-accepted non-gravitational
fields in physics. At the same time, energy conditions are strong enough to rule out
several EFE's solutions which are nonphysical in nature.
To study the physical properties of an anisotropic strange star completely,
we must have to check whether our model satisfies all the energy conditions
or not. The energy conditions, viz. Null Energy Condition (NEC), Weak Energy
Condition (WEC), Strong Energy Condition (SEC) and Dominant Energy Condition (DEC),
are satisfied if and only if the following inequalities hold simultaneously
at every point inside the fluid sphere:
\begin{eqnarray}
NEC&:& \rho \geq0,                 \label{eq37} \\
WEC&:& \rho+p_r \geq 0,~\rho+p_t \geq 0,  \label{eq38} \\
SEC&:& \rho+p_r+2p_t \geq 0,              \label{eq39} \\
DEC&:& \rho-|p_r| \geq 0,~\rho-|p_t| \geq 0.  \label{eq40}
\end{eqnarray}

Though, there are several matter distributions, for which SEC is violated from mathematical
prospect. According to Hawking~\cite{Hawking/1973}, any cosmological inflationary process
breaks SEC and for any scalar field along with a positive potential, this condition is not valid.

\begin{figure}[htp]
\centering
\includegraphics[width=4.4cm]{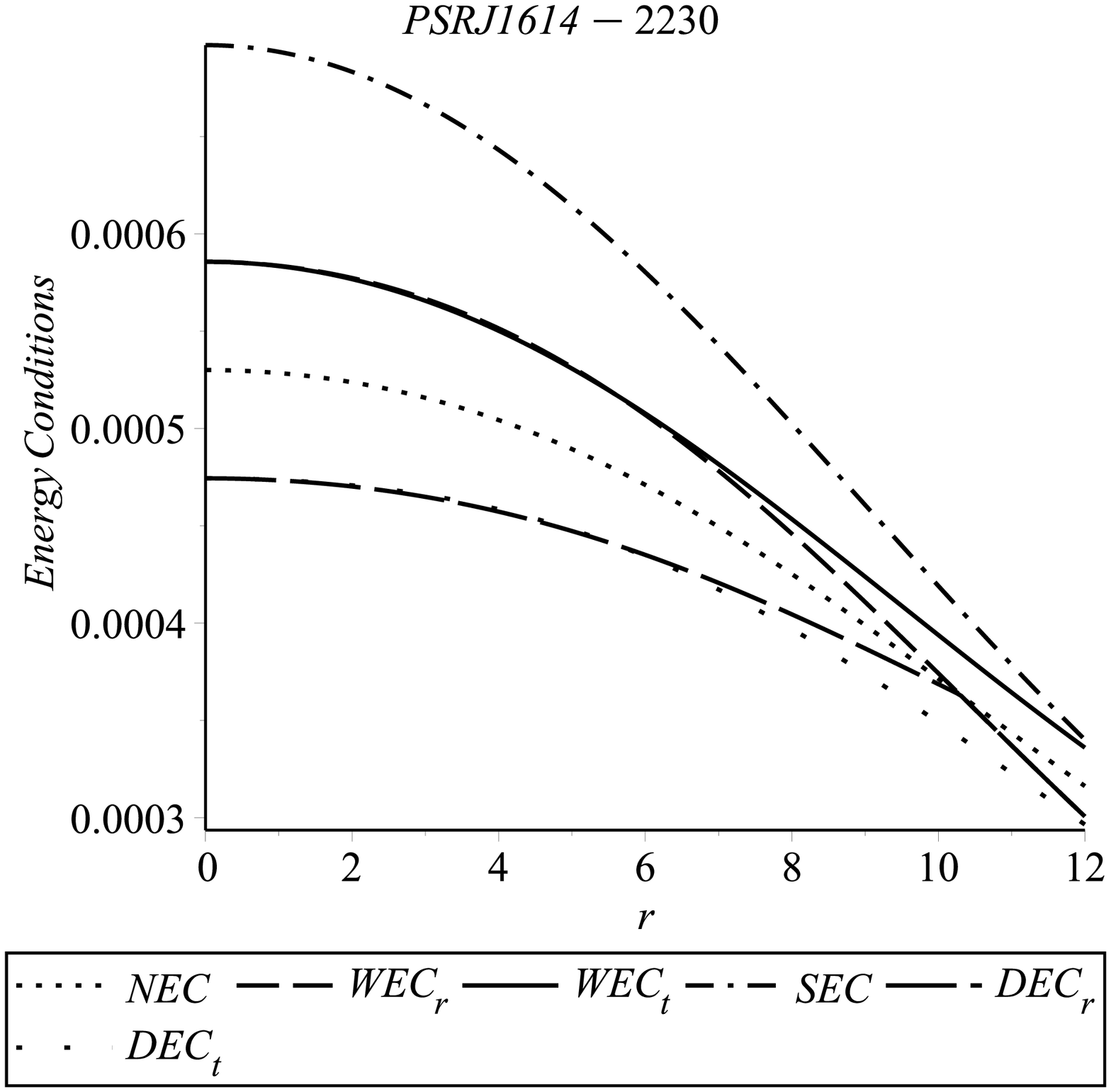}
\includegraphics[width=4.4cm]{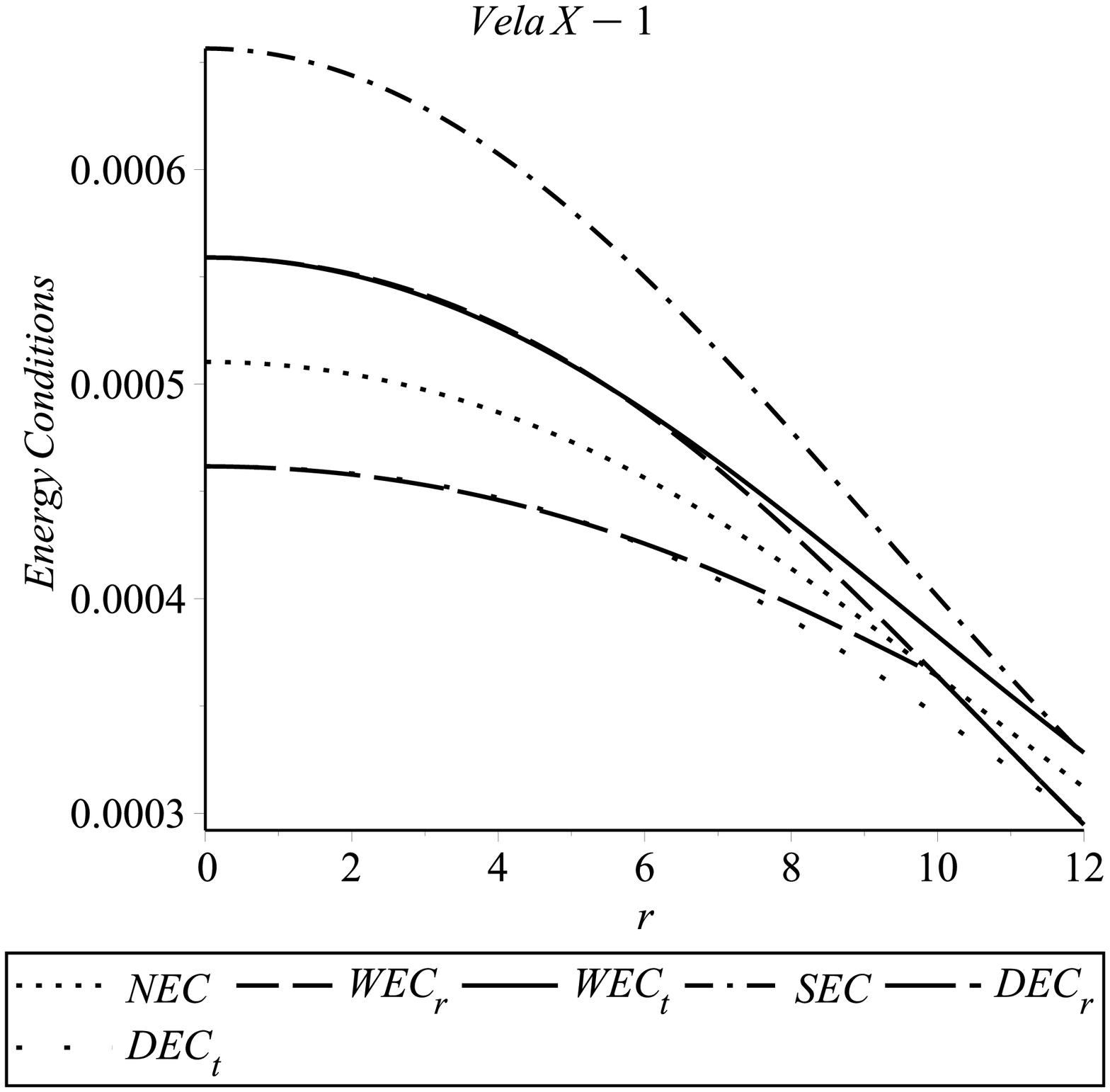}
\includegraphics[width=4.4cm]{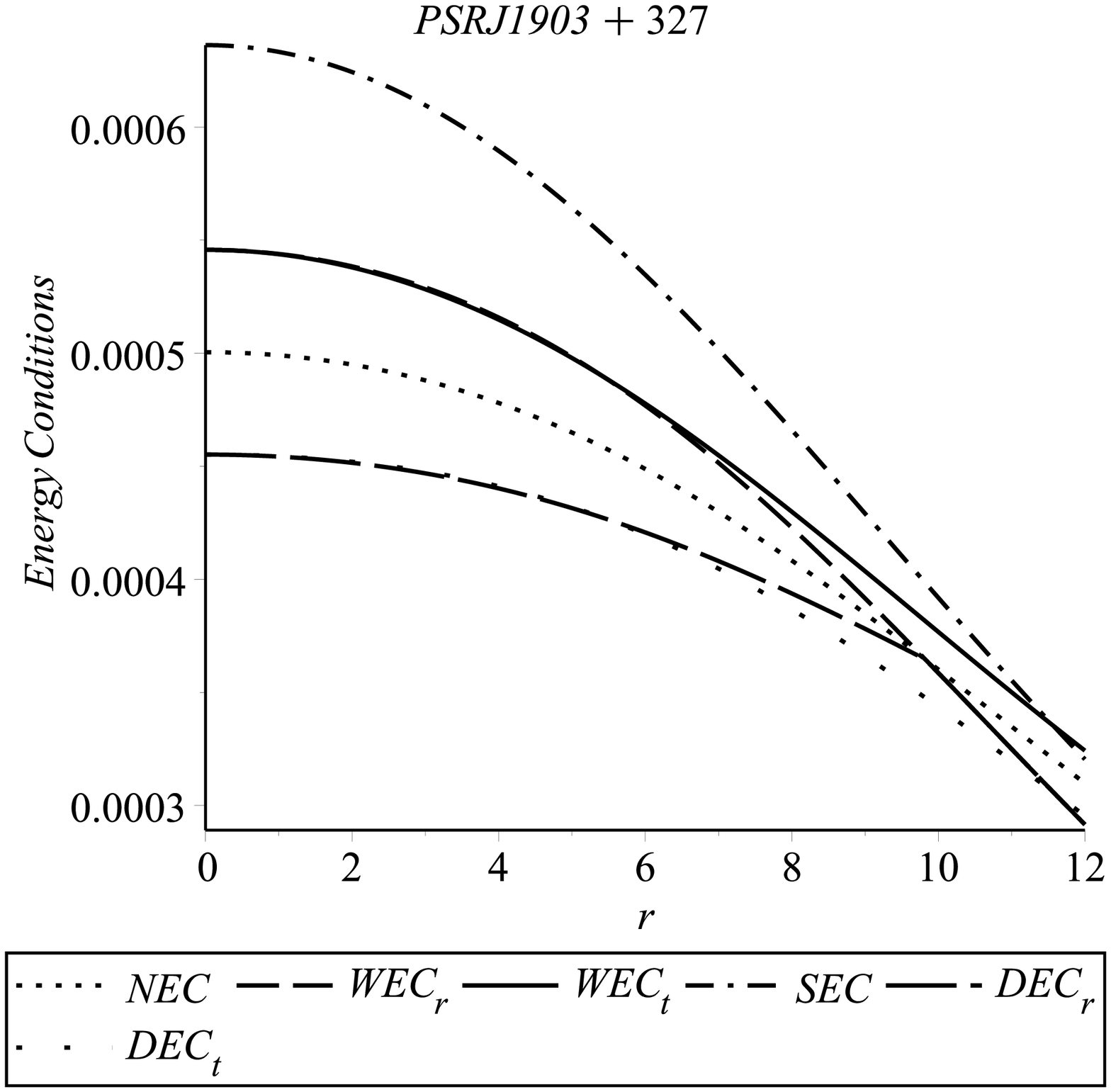}
\includegraphics[width=4.5cm]{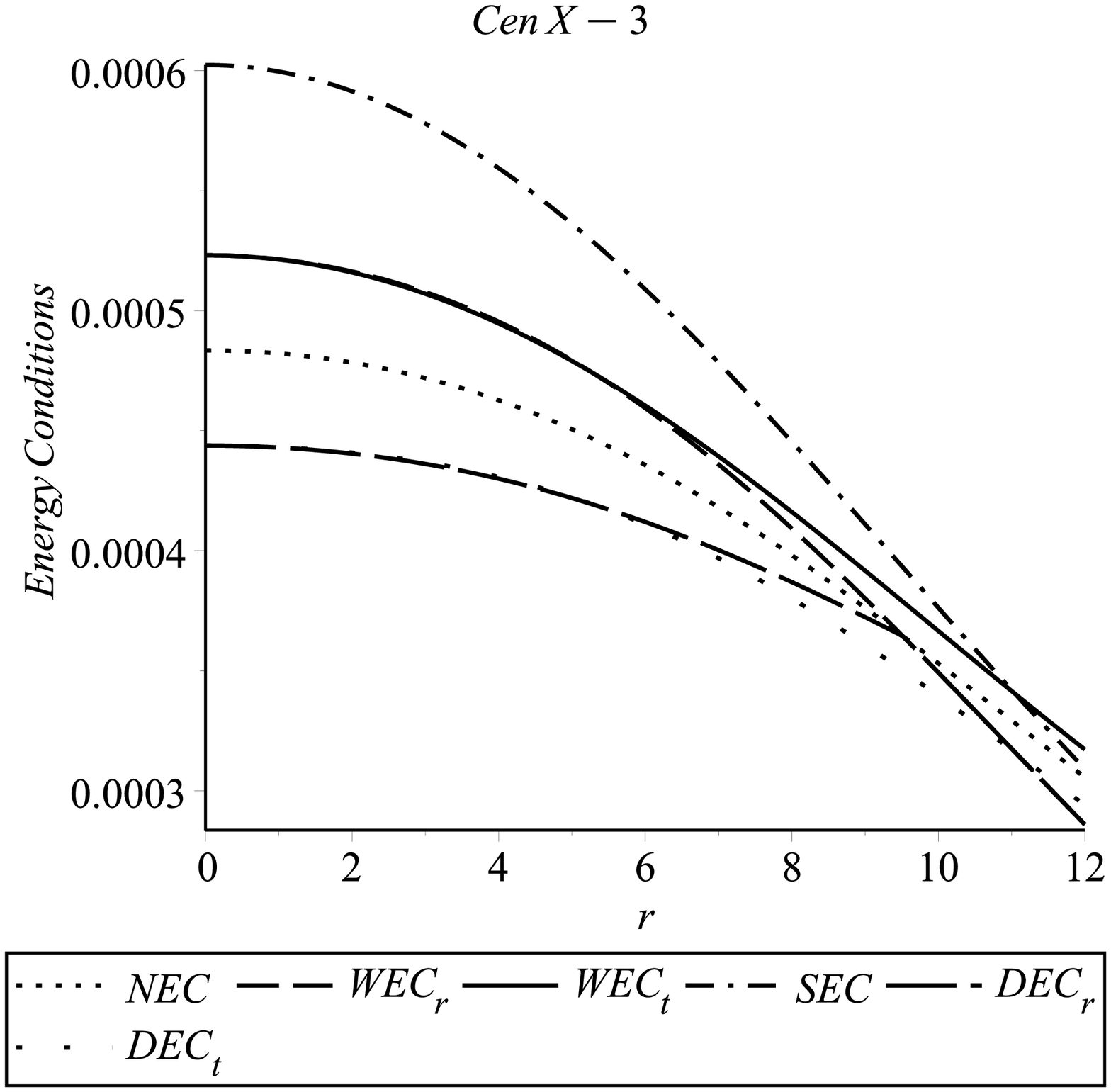}
\includegraphics[width=4.5cm]{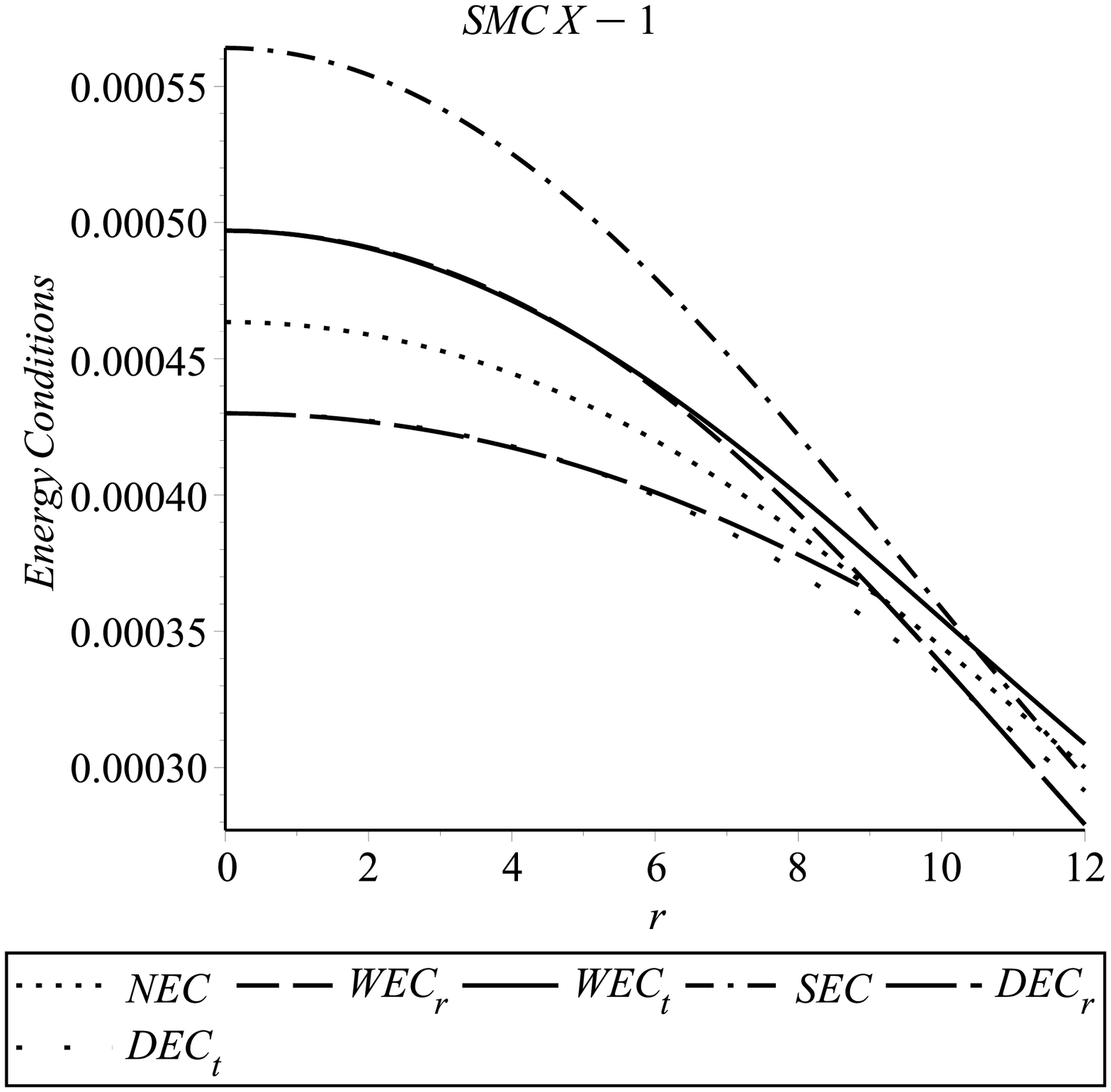}
\caption{Variation of the different energy conditions w.r.t. the radial
coordinate $r$ for different strange star candidates.}\label{ec.}
\end{figure}

Fig.~\ref{ec.} shows that our considered system with suitable choice of
mass and radius, is consistent with all the energy conditions and
confirms the physical validity of our model.

\subsection{Herrera's condition for stability analysis}
To study the stability of our stellar model, Herrera's causality condition
and cracking concept~\cite{herrera/1979,herrera/1992} are very much accepted over last three decade.
According to the causality condition, square of the tangential ($v_{ts}^{2}$) and radial ($v_{rs}^{2}$) speed
of sound should follow the restriction $0\leq{v_{rs}^{2}}\leq1$ and $0\leq{v_{ts}^{2}}\leq1$. Another
concept tells us that the region for which square of the transverse speed of sound is smaller than the square of the radial speed of sound, is a potentially stable ~\cite{herrera/1979,herrera/1992,chan/1993,abreu/2007,Andreasson/2009}. So, Herrera ~\cite{herrera/1992} and Andr\'{e}asson~\cite{Andreasson/2009} demand
$|v^2_{rs}-v^2_{ts}| \leq 1$ for stable matter distribution. This condition is known as `no cracking', i.e., potentially stable region.

\begin{eqnarray}
&\qquad\hspace{-2.5cm}v^2_{rs}=\frac{dp_r}{d\rho}=\frac{dp_r/dr}{d\rho/dr}=\frac{1}{3},  \label{eq41}\\
&\qquad\hspace{-3.3cm}v^2_{ts}=\frac{dp_t}{d\rho}=\frac{dp_t/dr}{d\rho/dr}=\frac{2}{3}\frac{B^2b^2r^8+B^2abr^6+3Babr^4-6b^2r^4-B^2ar^2+X}{2Bb^2r^6+3Babr^4+6b^2r^4+Ba^2r^2+Y},    \label{eq42}
\end{eqnarray}
where, $X=Ba^2r^2+8Bbr^2-6abr^2-B^2+3Ba-2a^2+2b$ and
$Y=2Bbr^2+6abr^2+Ba+2a^2-2b$.

\begin{eqnarray}
&\qquad\hspace{-1cm}\frac{dp_t}{d\rho}=\frac{dp_r}{d\rho}+\frac{d\Delta}{d\rho}=\frac{dp_r}{d\rho}+\frac{d\Delta/dr}{d\rho/dr}, ~~|v^2_{rs}-v^2_{ts}|=|\frac{d\Delta/dr}{d\rho/dr}|. \label{eq42a}
\end{eqnarray}

\begin{figure}[!htp]
\centering
\includegraphics[width=4.5cm]{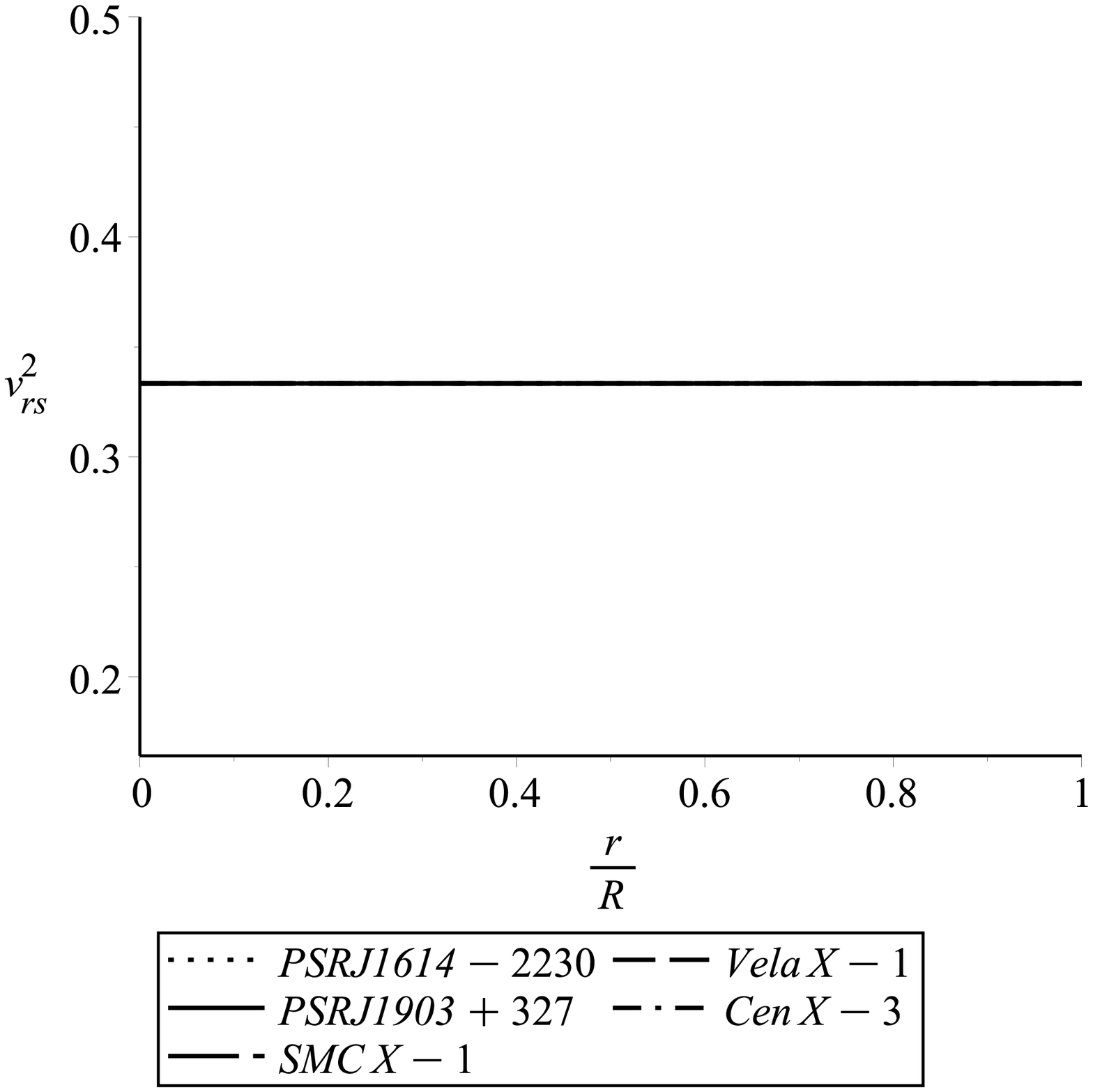}
\includegraphics[width=4.5cm]{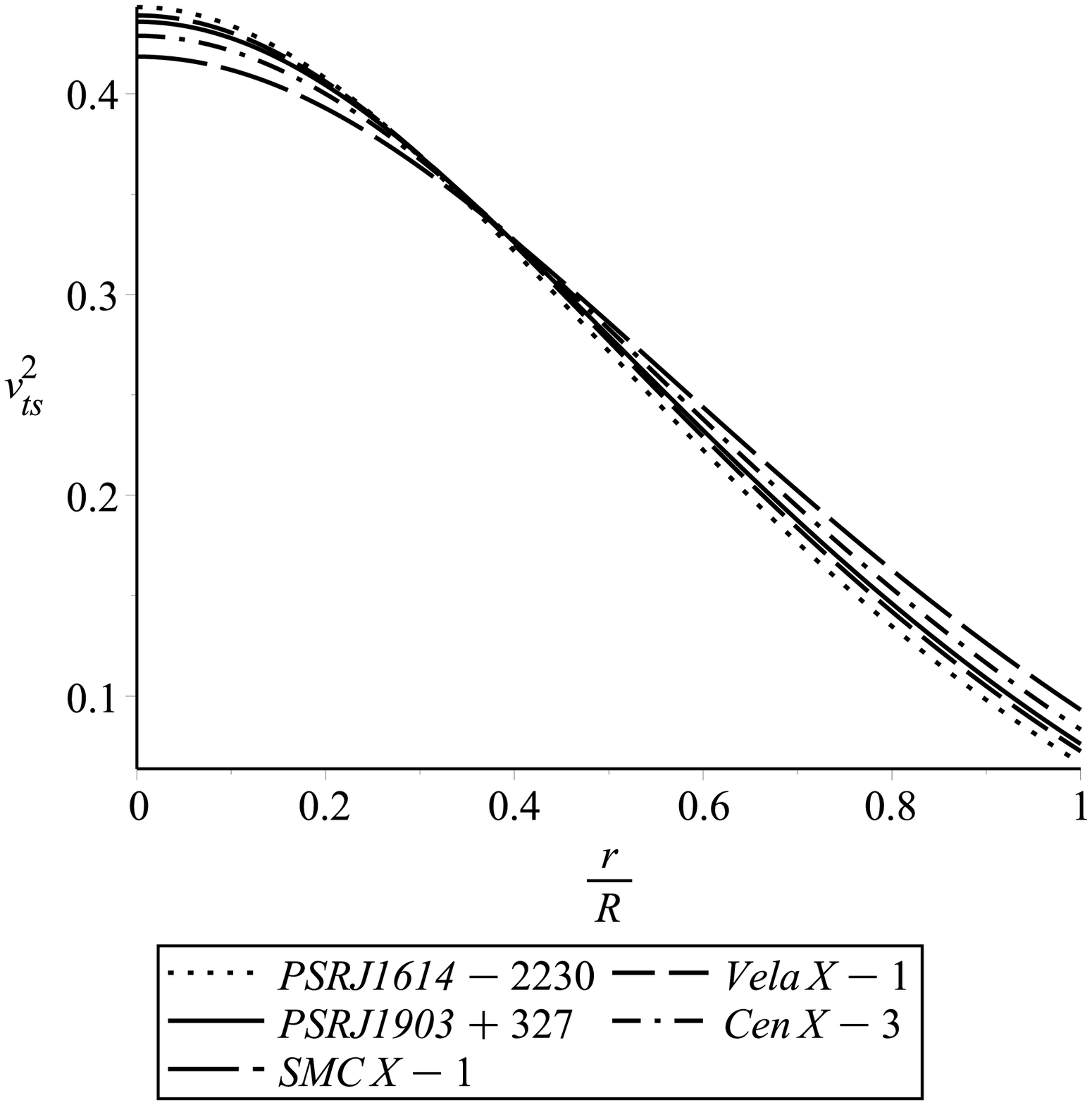}
\includegraphics[width=4.5cm]{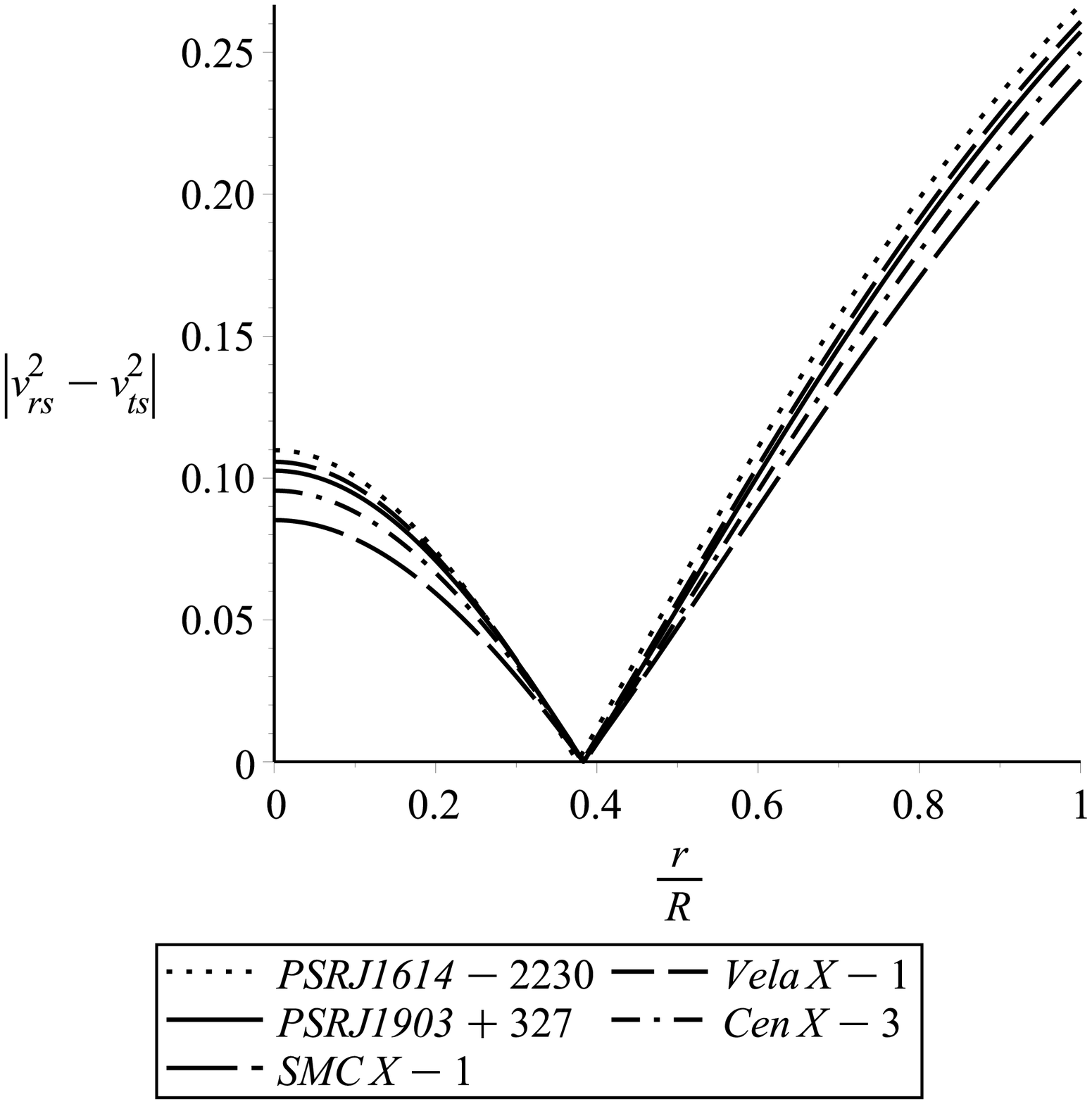}
\caption{Variation of $v^2_{rs}$ (left panel), $v^2_{ts}$ (right panel) and $|v^2_{rs}-v^2_{ts}|$ (lower panel) w.r.t.
the fractional radial coordinate $r/R$ for different strange star candidates.}\label{cau.}
\end{figure}

Variations of $v^2_{rs}$ and $v^2_{ts}$ (Eqs.~(\ref{eq41})-(\ref{eq42})) w.r.t. the fractional
radial coordinate $r/R$, have been displayed in Fig.~\ref{cau.}. Here graphs clearly show that
both $v^2_{rs}$ and $v^2_{ts}$ remain within their specified range $(0,~1)$ throughout the
stellar body. In Fig.~\ref{cau.}, magnitude of the difference between square of radial and
tangential sound speed, i.e., $|v^2_{rs}-v^2_{ts}|$ in Eq.~(\ref{eq42a}), first decreases with the
increasing radii and then bounces back to higher value, though always remain less than
unity. Hence our model maintains the consistency with Herrera's cracking concept and
re-establish the stable configuration.

\subsection{Adiabatic Index}
The term adiabatic index $\Gamma$, ratio of two specific heats~\cite{hillebrandt/1976}, characterizes
the stiffness of the EOS for a given density. We can use it to study the stability
of relativistic and non relativistic fluid sphere. According to Bondi~\cite{bondi/1964},
for an anisotropic fluid sphere, $\Gamma$ can be written as radial adiabatic index ($\Gamma_r$)
and tangential adiabatic index ($\Gamma_t$). It is used to study the dynamical stability
of the stellar system against an infinitesimal radial adiabatic perturbation
~\cite{chandrasekhar/1964,bardeen/1966,wald/1984,knutsen/1988,herrera/1997,horvat/2011,doneva/2012,mak/2013,silva/2015}.
For a stable Newtonian sphere, the adiabatic index $\Gamma>\frac{4}{3}$ and in case of neutral equilibrium,
$\Gamma=\frac{4}{3}$ according to Bondi~\cite{bondi/1964}. For isotropic sphere in relativistic
case, the above condition modifies due to the effect of regenerative pressure which leads
the sphere to become more unstable. But for general relativistic anisotropic sphere, more
complications arise because the nature of anisotropy decides the stability of the stellar system.
The adiabatic index should exceed $\frac{4}{3}$ inside a dynamically stable, relativistic,
anisotropic stellar system as predicted by Chan et al.~\cite{Chan/1993}, Heinzmann~\cite{heintzmann/1975}
and Hillebrandt~\cite{hillebrandt/1976}. For our proposed model the radial ($\Gamma_r$) and
 tangential ($\Gamma_t$) adiabatic indices can be defined as
\begin{eqnarray}
&\qquad\hspace{-0.5cm}\Gamma_r=\left[\frac{\rho(r)+p_r(r)}{p_r(r)}\right]\left[\frac{dp_r}{d\rho}\right]=\left[\frac{\rho(r)+p_r(r)}{p_r(r)}\right]\left[v_{rs}^2\right], \label{eq43}  \\
&\qquad\hspace{-0.5cm}\Gamma_t= \left[\frac{\rho(r)+p_t(r)}{p_t(r)}\right]\left[\frac{dp_t}{d\rho}\right]=\left[\frac{\rho(r)+p_t(r)}{p_t(r)}\right]\left[v_{ts}^2\right].  \label{eq44}
\end{eqnarray}

\begin{figure}[!htp]
\centering
\includegraphics[width=6cm]{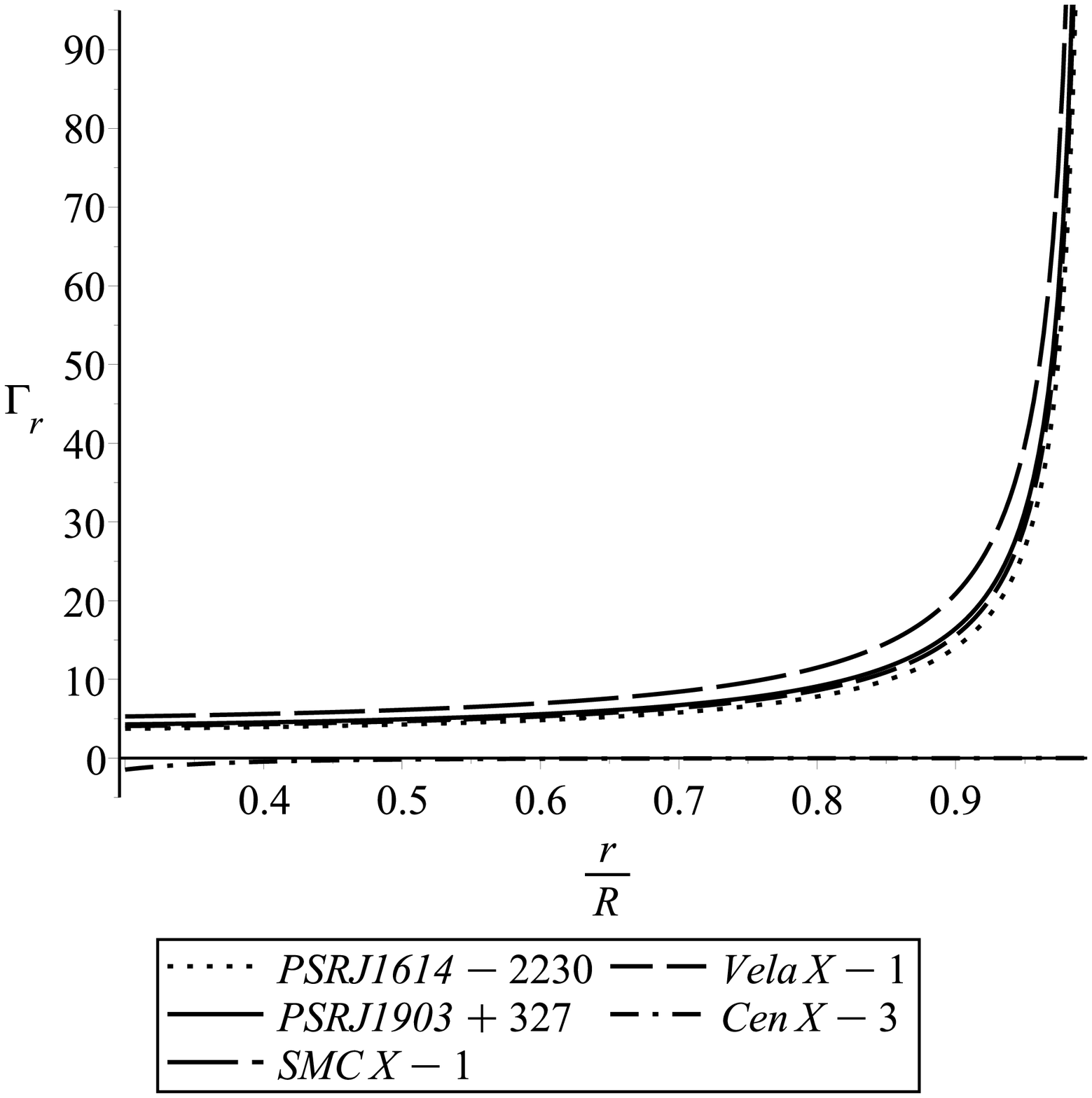}
\includegraphics[width=6cm]{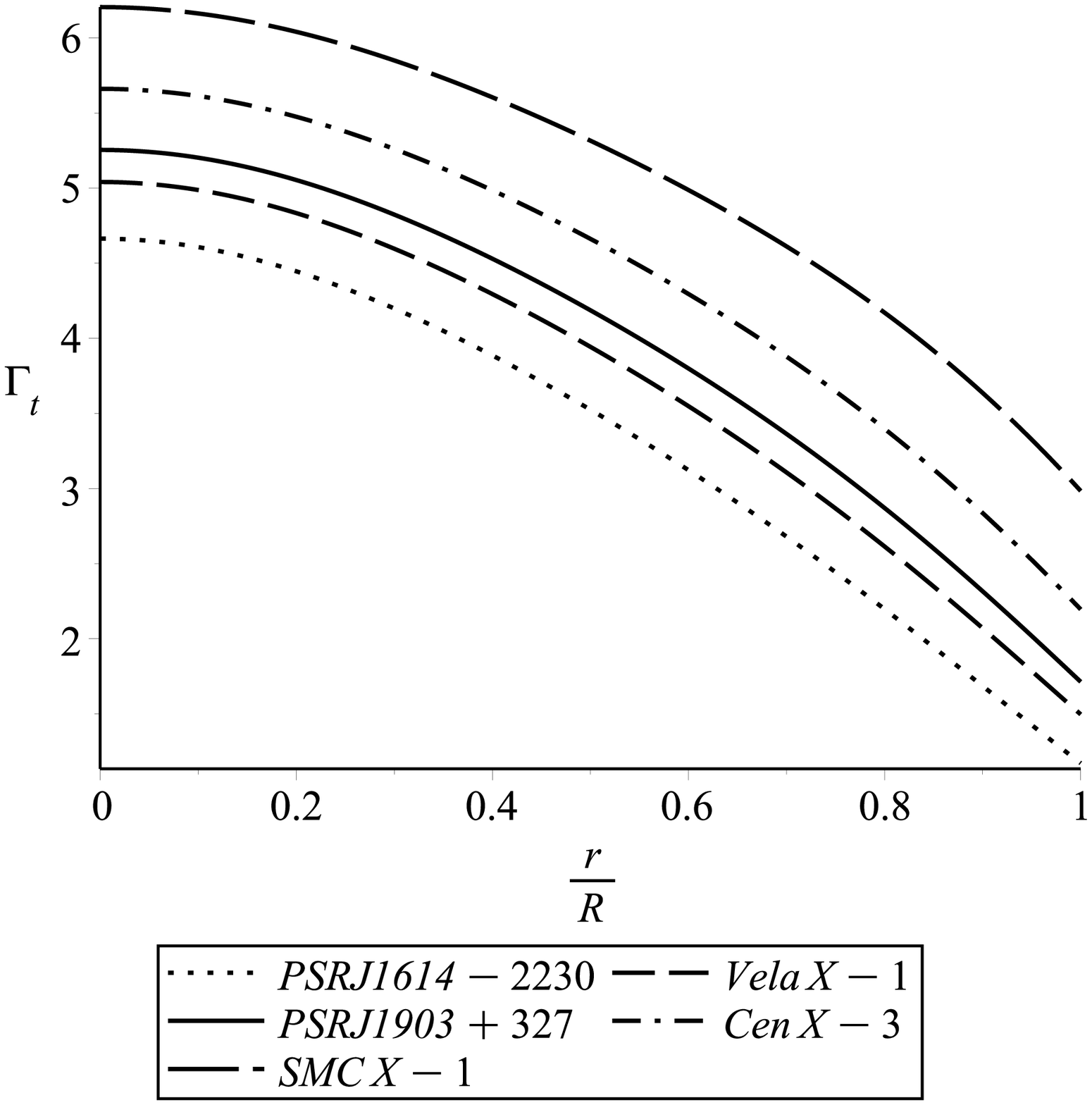}
\caption{Variation of adiabatic index $\Gamma_{r}$ (left panel) and $\Gamma_{t}$ (right panel)
 w.r.t. the fractional radial coordinate $r/R$ for different strange star candidates.}\label{adi.}
\end{figure}

We have shown the graphical representation of radial and tangential adiabatic indices
in Fig.~\ref{adi.}. One can see that the adiabatic indices are greater than $\frac{4}{3}$
through out the stellar system and it represents a stable configuration against the radial perturbation.

\subsection{Harrison-Zel$'$dovich-Novikov static stability criteria}
To examine the stability of stars, Chandrasekhar~\cite{Chandrasekhar/1964a}, Harrison et al.
~\cite{Harrison/1965} etc. calculate the eigen-frequencies for all the fundamental modes.
Later, following Harrison et al.~\cite{Harrison/1965}, Zel$'$dovich and Novikov
~\cite{Zeldovich/1971} simplified the calculations. To make it simpler, they have presumed
that adiabatic index for a pulsating star is comparable as for a slowly deformed matter.
This presumption leads to the criteria that mass of the star must be increasing in nature
with central density (i.e., $\frac{dM}{d\rho_c}>0$) to achieve stable configuration and
be unstable if $\frac{dM}{d\rho_c}<0$.

In our proposed model,  we can represent mass in terms of central density as follows
\begin{eqnarray}
M(\rho_c)=\frac{R(-3R^2b+16\pi B_g-16\pi\rho_c+3B)}{2(-3R^2b+16\pi B_g-16\pi\rho_c+3B-\frac{3}{R^2})}.\label{eq45}
\end{eqnarray}

Differentiating Eq.~(\ref{eq45}) w.r.t. $\rho_c$ we get,
\begin{eqnarray}
&\qquad\hspace{-.8cm}\frac{dM}{d\rho_c}=\frac{24\pi}{R\left(16\pi B_g-3R^2b-16\pi\rho_c+3B-\frac{3}{R^2}\right)^2}>0.\label{eq46}
\end{eqnarray}

\begin{figure}
\centering
\includegraphics[width=6cm]{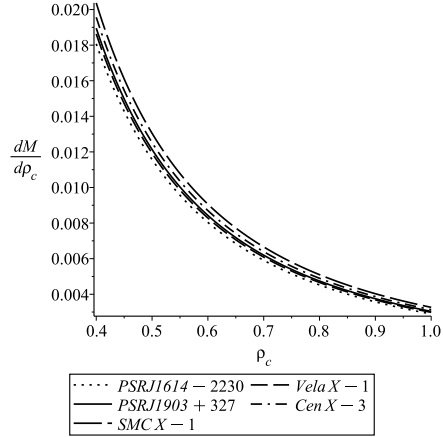}
\caption{Variation of $\frac{dM}{d\rho_c}$ w.r.t. $\rho_c$ for different strange star candidates.}\label{HZN.}
\end{figure}

Fig.~\ref{HZN.} also shows that $\frac{dM}{d\rho_c}$ is always positive throughout
the stellar model. So, the Harrison-Zel$'$dovich-Novikov condition~\cite{shee/2016,Bhar/2017}
further confirms  the stability of our proposed model.

\section{Mass-Radius relation and surface redshift}
\begin{figure}[!htp]
\centering
\includegraphics[width=4.4cm]{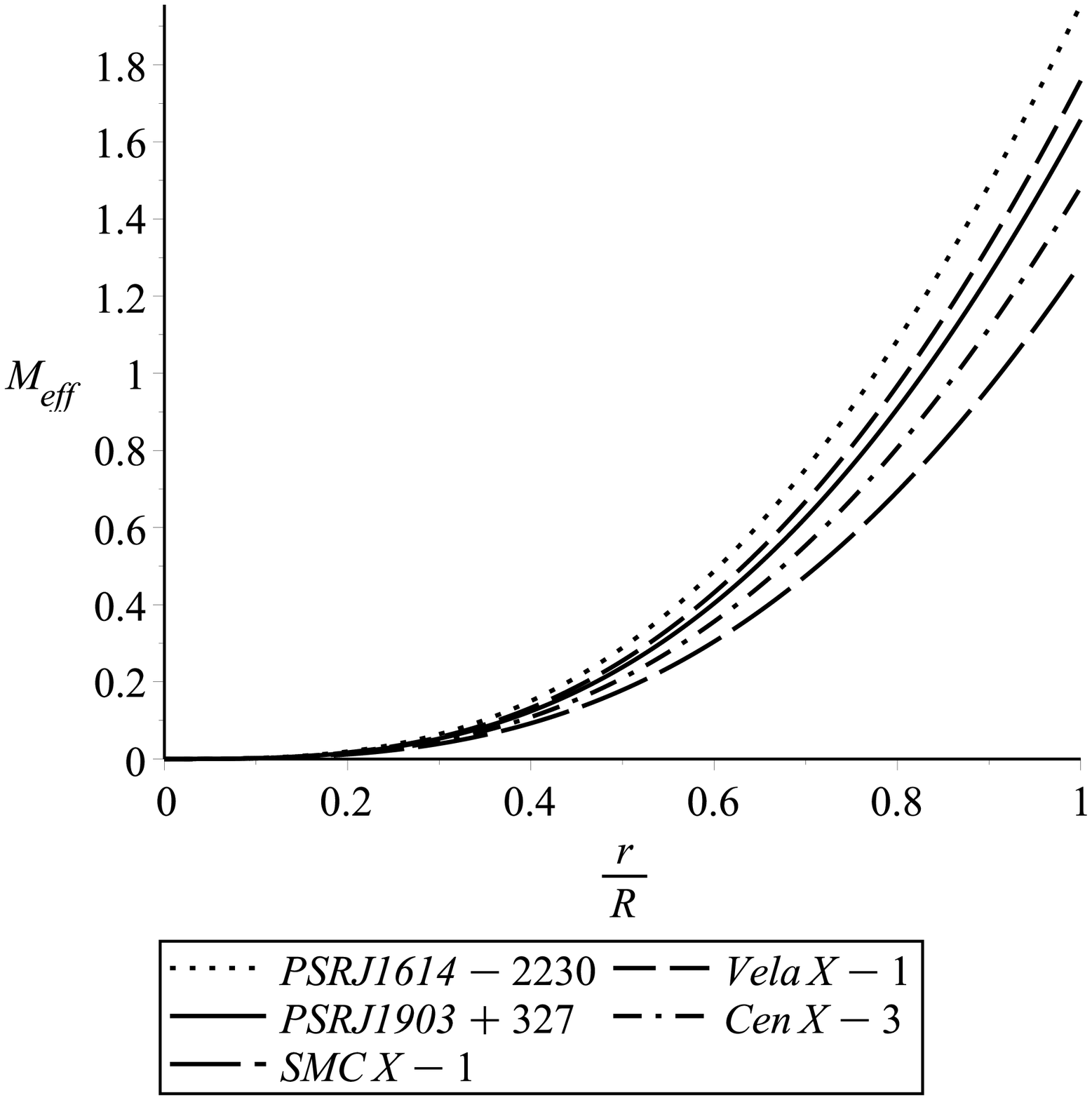}
\includegraphics[width=4.4cm]{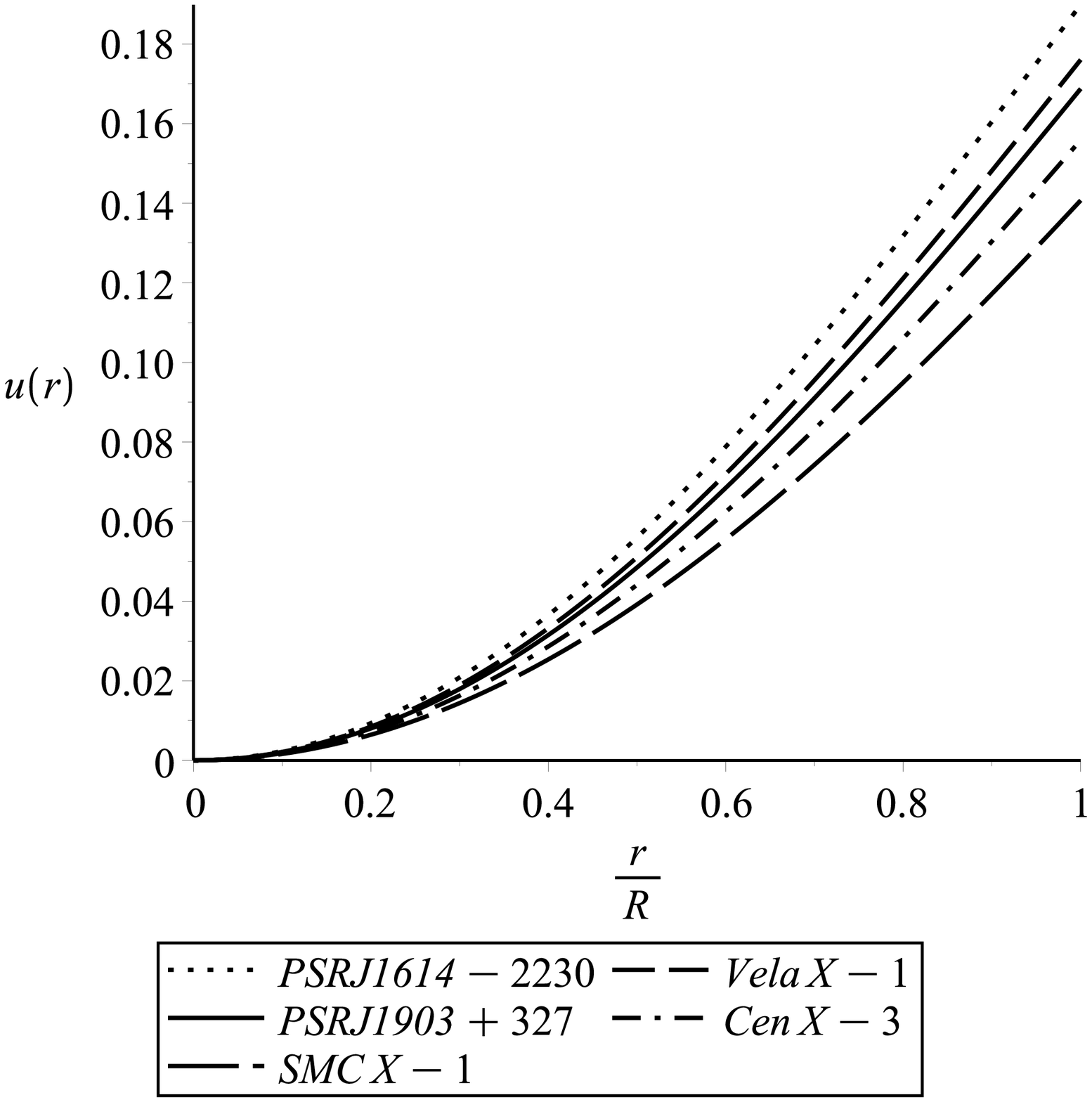}
\includegraphics[width=4.4cm]{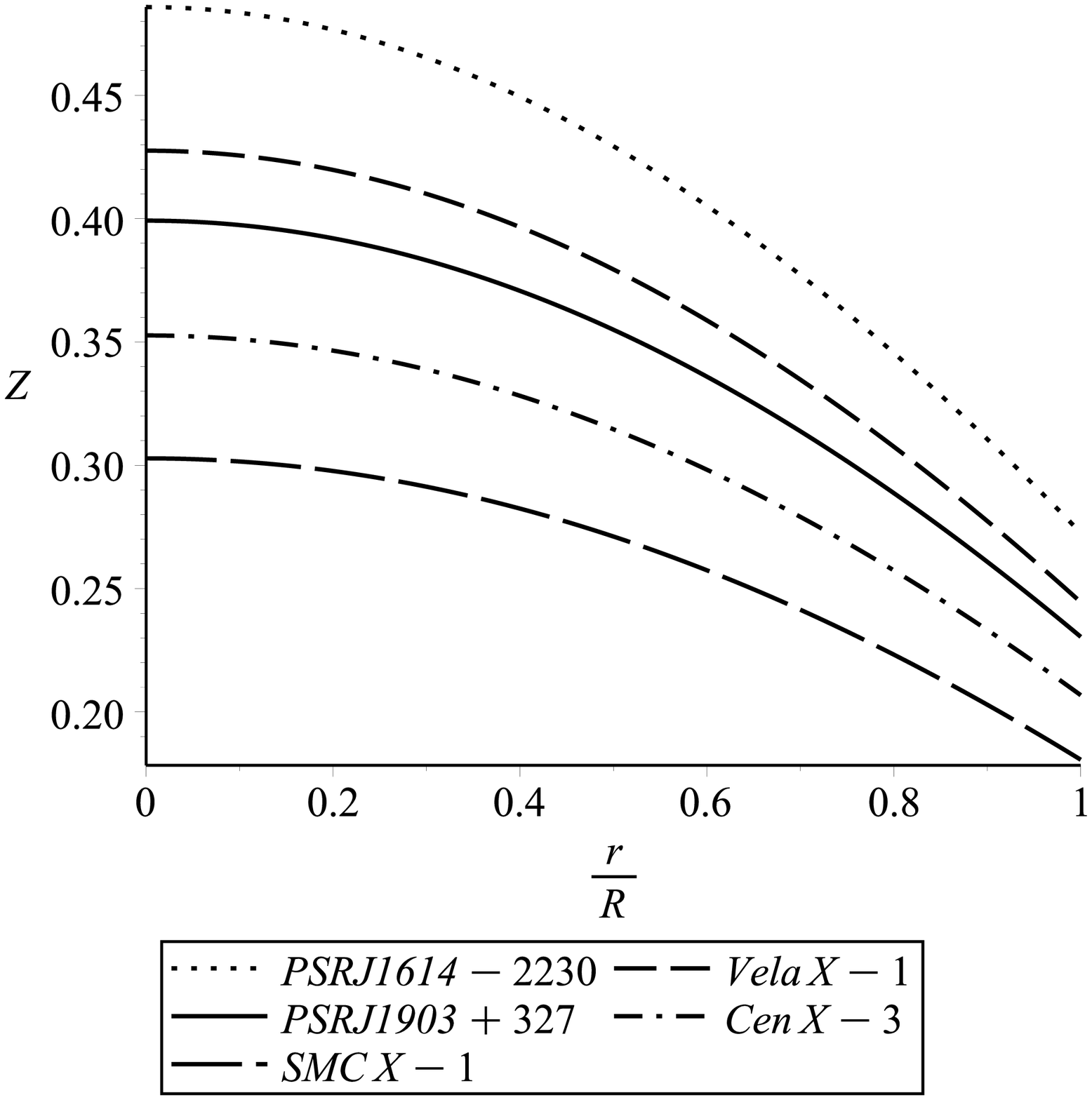}
\caption{Variation of effective mass (left panel), compactification factor (middle panel) and gravitational redshift (right panel) w.r.t. the fractional radial coordinate $r/R$ for
different strange star candidates.}\label{sur.}
\end{figure}

For a static, spherically symmetric, anisotropic fluid distribution we can calculate its
effective mass from the density profile, as given in Eq.~(\ref{eq10}) and represent it graphically
in Fig.~\ref{sur.}. Hence, effective mass is
\begin{eqnarray}
M_{eff} &=& \int _{0}^{R}4\pi r^2\rho(r)dr. \label{eq47}
\end{eqnarray}

From Fig.~\ref{sur.}, it is clear that as $r \rightarrow 0$, $M_{eff} \rightarrow 0$,
so the effective mass is a monotonic increasing function.

For a static spherically symmetric perfect fluid star, Buchdahl~\cite{buchdahl/1959} derived
an upper limit for maximum allowed mass to radius ratio, i.e., $\frac{2M}{R}<\frac{8}{9}$.
Mak et al.~\cite{mak/2001} generalised the result for charged sphere. The term $\frac{M}{R}$
is known as compactness which classifies the stellar objects into different categories as
follows~\cite{jotania/2006}, (i) normal star: $M/R\sim 10^{-5}$, (ii) white dwarf:
$M/R\sim 10^{-3}$, (iii) neutron star: $10^{-1}<M/R<1/4$, (iv) ultra dense compact
star: $1/4<M/R<1/2$ and (v) black hole: $M/R=1/2$.

The compactness $u(r)$ for our model is given by
\begin{eqnarray}
u(r)=\frac{M_{eff}}{r}&=& \frac{1}{r}\int _{0}^{R}4\pi r^2\rho(r)dr. \label{eq48}
\end{eqnarray}

The graphical representation of compactness (Fig.~\ref{sur.}) shows that it monotonically
increases with the radius of the star and it's maximum value $>0.25$, implies that our
model corresponds to an ultra dense star.

The surface redshift $(Z_s)$ and gravitational redshift $(Z)$ for our model can be represented by the following relations
\begin{eqnarray}
Z_s&=&\frac{1}{\sqrt{1-2u(r)}}-1, \label{eq49} \\
Z&=&e^{\frac{-\nu(r)}{2}}-1. \label{eq50}
\end{eqnarray}

In absence of cosmological constant, Barraco and Hamity~\cite{Barraco/2002} proved that
$Z_s\leq2$ for an isotropic star. Later Bohmer and Harko~\cite{Bohmer/2006} established
that or an anisotropic star, surface redshift can reach much higher value $Z_s\leq5$,
in presence of cosmological constant. Though the restriction was eventually modified
and maximum acceptable value was calculated  as $Z_s=5.211$ ~\cite{Ivanov/2002}. In our case, we get $Z_s\leq1$
for different strange star candidates (Table 2). The variation of the gravitational redshift w.r.t.
$r/R$ shown in Fig.~\ref{sur.}, decreases monotonically with increasing radii. $Z$ is positive
as well as finite throughout the system, which strongly supports the acceptance of relativistic
strange star model.

\section{Discussion and Conclusion}

\begin{table}[!htp]
\centering \caption{Mass and radius of different strange star~\cite{Rahaman/2014} candidates}\resizebox{\columnwidth}{!}{
      \begin{tabular}{@{}llllllll@{}}
\hline Case & Stars                     & Mass        & Radius & $\frac{M}{R}$      & $B_g$   & $Z_s$   \\
                  &                     &(km)       &(km)&                 & {\small{(MeV/{fm}$^3$)}}    \\
\hline        I   & $PSR~J~1614~2230$   & 1.97        & 10.3       & 0.191          & 68.865  & 0.27    \\
              II  & $Vela~X-1$          & 1.77        & 9.99       & 0.177          & 69.097  & 0.24    \\
              III & $PSR~J~1903+327$    & 1.667       & 9.82       & 0.17           & 69.161  & 0.23    \\
              IV  & $Cen~X-3$           & 1.49        & 9.51       & 0.157          & 69.144  & 0.21    \\
              V   & $SMC~X-1$           & 1.29        & 9.13       & 0.1413         & 68.84   & 0.18    \\
\hline \label{Table1}
\end{tabular}}
\end{table}

\begin{table}[!htp]
\centering \caption{Determination of model parameters $a$, $b$, $C$, $B$, central density,
surface density, radial pressure for different strange star candidates}\resizebox{\columnwidth}{!}{
      \begin{tabular}{ccccccccc}
\hline Case  & $a$       & $b$        & $B$      &$C$   & $\rho(r=0)$ & $\rho(r=R)$   & $p_r(r=0)$     \\
         &{\small{({km}$^{-2}$)}}&{(km$^{-4}$)}&{\small{({km}$^{-2}$)}}& &{\small{({gm/cm}$^3$)}}&{\small{({gm/cm}$^3$)}} &{\small{({dyne/cm}$^2$)}}  \\
\hline   I     & 0.004441  &0.0000132  &0.00292  &0.6731  &$7.1713\times10^{14}$ &$4.912\times 10^{14}$  &$6.753\times 10^{34}$ \\
         II    & 0.004276  &0.0000123  &0.00275  &0.7005  &$6.905\times10^{14}$  &$4.929\times 10^{14}$  &$5.91\times 10^{34}$  \\
         III   & 0.004193  &0.0000118  &0.00267  &0.7147  &$6.771\times10^{14}$  &$4.933\times 10^{14}$  &$5.493\times 10^{34}$ \\
         IV    & 0.00405   &0.000011   &0.002523 &0.7393  &$6.541\times10^{14}$  &$4.932\times 10^{14}$  &$4.81\times 10^{34}$ \\
         V     & 0.003883  &0.0000101  &0.00236  &0.7676  &$6.271\times10^{14}$  &$4.91\times 10^{14}$   &$4.067\times 10^{34}$  \\
\hline \label{Table2}
\end{tabular}}
\end{table}

In the present article our motive is to explore the relevance of TK metric potentials in the modeling of strange quark stars. Considering the SQM distribution, expressed as the
simplified MIT bag model EOS (Eq.~(\ref{eq8})) we have studied the strange star more prominently.
We have used Einstein's spacetime to construct the preliminary mathematical background.
From Einstein's field equations, Eqs.~(\ref{eq3})-(\ref{eq5}), by using the above
mentioned TK metric potentials (Eqs.~(\ref{eq14}) and (\ref{eq15})) and MIT bag EOS (Eq.~(\ref{eq8})),
we have found out the solutions of various physical interest and also shown their variation
graphically w.r.t. the radial distance $r$.

The metric potentials also specify the geometry of the space time. Their
graphical representations (FIG.~\ref{pot.}) tell us that both the metric
potentials satisfy the necessary condition, i.e., $e^{\nu(r)}(r=0)$ is non-zero
positive and $e^{\lambda(r)}(r=0)=1$, so that the solutions should be free from
physical and geometrical singularities. Both $e^{\nu(r)})$ and $e^{\lambda(r)}$ increase
monotonically and non-linearly from centre to the surface for all the stars.

However in the framework of GR, our calculated value for $B_g$ remains in the range
$(55-75)$~MeV/fm$^{3}$, for stable SQM distribution~\cite{Farhi1984,Alcock1986,weber/2005}, e.g. for star
$PSRJ$ $1614~2230$ ($1.97~M_\odot$ and $R=10.3$~km) $B_g=68.865$~MeV/fm$^{3}$, for
star $PSR~J~1903+327~(1.667~M_\odot$ and $R=9.82$~km) $B_g=69.161$~MeV/fm$^{3}$ and for
a relatively lower mass ($1.29~M_\odot)$ star $SMC~X-1$ with radius $R=9.13$~km, $B_g=68.84$~MeV/fm$^{3}$.
Hence it is fascinating to mention that our calculated $B_g$ values satisfy not only the theoretical stability criteria
~\cite{Farhi1984,Alcock1986,weber/2005} but also exactly match to the observational result of RHIC
and CERN-SPS~\cite{burgio/2002}. Our study may be generalised by considering more realistic EOS
obtained from QCD simulations~\cite{alford/2013}.

All the figures clearly establish the non-singular, stable nature from both
geometrical, physical aspects. All the calculated values for different physical
parameter are provided in Tables~(\ref{Table1} and~\ref{Table2}). Here, the important
results from this strange star model have been discussed as follows:

$\textbf{1. Density and Pressure:}$
Solving EFE (Eqs.~(\ref{eq3})-(\ref{eq5})) we get the density ($\rho$), radial
pressure ($p_r$) as well as tangential pressure ($p_t$) as shown in Eqs.~(\ref{eq10})-(\ref{eq12})
respectively. Central matter density $\rho_c$ of Eq.~(\ref{eq16}) and central radial pressure ($p_{rc}$)
can also be calculated by putting $r=0$ in Eqs.~(\ref{eq10}) and (\ref{eq11}). Density profile~
($\rho$) and both the pressures ($p_r$ and $p_t$) are free from central singularity.
At the centre of the star $\rho_c$, $p_r$ and $p_t$ are finite and then decreases continuously
with the increasing radius as shown in Fig.~\ref{pres.}. In Table~(\ref{Table2})
we have provided the values of density at the centre as well as at the surface for different strange stars.
These numerical values give a clear idea that central density is higher than the surface density which
is expected for an ultra dense star made of strange quark~\cite{ruderman/1972,Glendenning/1997,Herzog/2011}.
For example, in case of $PSR~J~1614~2230$, $\rho_c\sim7.1713\times10^{14}$
~{gm/cm}$^3$ which reduces upto $68\%$ at the surface, where density is $4.912\times 10^{14}$~{gm/cm}$^3$.
Radial pressure at the centre of the star also very high, i.e., $(4.07-6.75)\times10^{34}$~dyne/cm$^2$.
According to Fig.~\ref{pres.}, our model shows the spheroidal nature~\cite{quevedo/1989,chifu/2012,shee/2017}.
Variation of anisotropic stress w.r.t. the fractional radial coordinate $r/R$ in Fig.~\ref{pres.} also
establish the stable nature of our proposed model.

$\textbf{2. TOV equation:}$
In our proposed model, generalised TOV Eq.~(\ref{eq32}) satisfies the stability criteria~\cite{Leon/1987}.
Fig.~\ref{tov.} clearly shows that combined effect of hydrostatic force ($F_h$) and anisotropic
force ($F_a$) balances the effect of gravitational force ($F_g$). For every strange star
candidates, we get almost same nature of force.

$\textbf{3. Energy conditions:}$
Another strong reason behind the acceptability of our proposed model is that our model
satisfies all the energy conditions, viz. NEC, WEC, SEC and DEC as mentioned in
Eqs.~(\ref{eq37})-(\ref{eq40}). Graphical representations of these conditions, shown in
Fig.~\ref{ec.} tell that all the energy values are maximum at the centre but reduces
gradually towards the surface.

$\textbf{4. Stabilty of model:}$
According to Herrera's~\cite{herrera/1992} cracking concept, $v^2_{rs}$ and $v^2_{ts}$ should remain between
the range $0$ and $1$. Here, following MIT bag model [Eq.~(\ref{eq8})], $v^2_{rs}$ maintain a
constant value $\frac{1}{3}$ throughout the stellar body and $v^2_{ts}$ in Eq.~(\ref{eq42}),
always remain in the above  mentioned range, which is clearly shown in Fig.~\ref{cau.}. From
graph~(Fig.~\ref{cau.}), $v^2_{ts}$ becomes maximum ($\approx0.45$) at the centre and monotonically
reduces towards the surface but always remains positive. For a potentially stable
configuration, Herrera~\cite{herrera/1992} and Andr\'{e}asson~\cite{Andreasson/2009}
demand `no cracking' condition i.e., $|v^2_{rs}-v^2_{ts}|\leq1$  to be obeyed
and it is very clear from Fig.~\ref{cau.} that condition is also satisfied.

Both adiabatic index as defined in Eqs.~(\ref{eq43}) and (\ref{eq44}) have been plotted in Fig.~\ref{adi.}.
From their variation it's clear that both $\Gamma_r$ and $\Gamma_t\geq\frac{4}{3}$ at the centre
and maintain Bondi's~\cite{bondi/1964} criteria for static configuration throughout the total
stellar body.

Another stability criteria is that, value of EOS parameter should lie in the range
$0-\frac{1}{3}$. Here, variation of both $\omega_r$ and $\omega_t$ have been
featured graphically in Fig.~\ref{eos.}. From figure it's clear that both $\omega_r$, $\omega_t$
are positive through out the star and obtain maximum value ($\approx 0.11$) at the centre and
reduces towards the surface area. Hence the fluid distribution is non-exotic~\cite{shee/2016} in nature.

Harrison-Zel$'$dovich-Novikov~\cite{Harrison/1965,Zeldovich/1971} static stability criteria is also satisfied [Eq.~(\ref{eq46})]
for different strange stars in our proposed model. Fig.~\ref{HZN.} undeniably shows that variation of
$\frac{dM}{d\rho_c}$ is always remain positive within the entire stellar body and
reduces towards the surface from maximum value at the centre similar as EOS parameter.

$\textbf{5. Buchdahl Condition:}$
Fig.~\ref{sur.} shows the variation of effective mass function ($M_{eff}$) w.r.t. the fractional radial coordinate $r/R$.
Variation shows that at $r\rightarrow0$, $M_{eff}\rightarrow0$, i.e., mass function is regular at the centre.
According to Buchdahl~\cite{buchdahl/1959} condition, mass radius ratio, i.e., $\frac{M}{R}\leq\frac{4}{3}$,
for static, spherically symmetric and perfect fluid distribution. In the present study, we
have considered $5$ different strange star candidates (Table~\ref{Table1}) with their observed
radius and mass value, for which Buchdahl~\cite{buchdahl/1959}
condition is perfectly obeyed and the ratio remains in the range ($0.14-0.19$).

$\textbf{6. Compactness and surface redshift:}$
In this model, variation of compactness ($u_r$) and gravitational redshift ($Z_s$) have been displayed in Figs.~\ref{sur.}. 
From figures it's clear that both the functions are continuous at $r\rightarrow0$
and increases for higher $r$ value for different strange stars. Calculated values for the surface redshift
become maximum for star $PSR~J~1614~2230$ $(Z_s=0.27)$ and $Z_s=0.18$ for relatively low mass and
smaller radius star $SMC~X-1$. Here the $Z_s$ value strongly establish that our discussed candidates
must be strange stars.

From literature survey, it's obvious that various group of researchers have studied
strange stars in the framework of GR. But, few of them have introduced such a model
that can't satisfy all the stability conditions at a time. Again few proposed models
show the stability throughout the stellar system, but singularity arises at the centre.
Few works do not satisfy all the stability criteria, energy conditions, Buchdahl limit~\cite{buchdahl/1959}
one at a time. However in our current study, using the TK metric and the EOS for MIT
bag model under the framework of GR, we get a such a model that can overcome all the
stability issues as well as exhibit the non-singular nature throughout the stellar body.

It is already mentioned, in a sketchy way, in Introduction that recently Jasim et al.~\cite{jasim/2018}
have studied anisotropic strange star using TK metric ~\cite{tolman/1939,kuchowicz/1968}.
In their study, they have assumed the value of bag constant as a `fixed quantity', i.e., $60$~MeV/fm$^{3}$
for all the strange star candidates under consideration with massive quark condition. As `bag constant' is
an inherent characteristics of strange stars so it can not be exactly same for different strange stars and
hence the study by Jasim et al.~\cite{jasim/2018} seems lacking for physical acceptance. However, from our model
we can calculate the bag values for individual strange star, knowing their observed mass and radius only
under the massless quark condition which fall within the proposed range $(55-75)$~MeV/fm$^3$~\cite{Farhi1984,Alcock1986}.
Besides, our study provides more simplified solutions for $\rho$, $p_r$ and $p_t$, using EFEs and MIT bag EOS.
Jasim et al.~\cite{jasim/2018} also have introduced the cosmological constant $\Lambda$ which is
zero in our case. Now, as a comparative study, we can observe that by setting only $\Lambda=0$
in~\cite{jasim/2018}, one can not simply retrieve our results which are quite different from
the results of Jasim et al.~\cite{jasim/2018}. An overall observation on both the works reveals
that the present investigation provides more general analysis of strange stars.

Hence an overall and final comment is that our proposed model is a representative of
singularity-free, stable and viable one which represents a highly compact star made of SQM
and perfectly fits for strange star candidates to analyze their different physically features.

\section*{Acknowledgement}
SR and FR are thankful to the Inter University Centre for Astronomy and Astrophysics (IUCAA) for providing Visiting Associateship under which a part of this work has been carried out. SR also thanks the Centre for Theoretical Studies (CTS), IIT Kharagpur, India for providing short term visit under which rest of the work has been completed. SB is thankful to DST-INSPIRE (INDIA) [IF~160526] for financial support and all type of facilities for continuing research work. SB and DS acknowledge their gratitude to research scholar Shounak Ghosh for his valuable suggestions which became fruitful in preparation of the manuscript.

\end{document}